\documentclass[useAMS,usenatbib]{mn2e}

\usepackage{color}
\usepackage{graphicx}
\usepackage{rotating}
\usepackage{longtable}
\usepackage{amsmath}
\usepackage{amstext}
\usepackage{amssymb}
\def   \aj {{\rm {AJ}}}
\def   \araa {{\rm {ARA\&A}}}
\def   \apj {{\rm {ApJ}}}

\def   \apjs {{\rm {ApJS}}}
\def   \apss {{\rm {Ap\&SS}}}
\def   \aap {{\rm {A\&A}}}

\def   \mnras {{\rm {MNRAS}}}

\def   \apjl {{\rm {ApJL}}}
\def   \nat {{\rm {Nature}}}

\title[Massive molecular outflows]{A distance limited sample of massive molecular outflows\thanks{drawn from the RMS survey, http://rms.leeds.ac.uk/cgi-bin/public/RMS\_DATABASE.cgi}}
\author[L. T. Maud et al.]
{L. T. Maud$^{1,2}$\thanks{E-mail:maud@strw.leidenuniv.nl (LTM)}, T. J. T. Moore$^{3}$, S. L. Lumsden$^{1}$, J. C. Mottram$^{2}$,
\newauthor J. S. Urquhart$^{4}$ and M. G. Hoare$^{1}$\\
$^{1}$School of Physics and Astronomy, University of Leeds, Leeds, LS2 9JT, UK\\
$^{2}$Leiden Observatory, Leiden University, PO Box 9513, 2300 RA Leiden, The Netherlands\\
$^{3}$Astrophysics Research Institute, Liverpool John Moores University, 146 Brownlow Hill, Liverpool, L3 5RF, UK\\
$^{4}$Max-Planck-Institute f\"{u}r Radioastronomie, Auf dem H\"{u}gel 69, D-53121 Bonn, Germany\\}
\begin{document}

\date{Accepted 2015 July 16. Received 2015 July 16; in original form 2015 Februray 23}

\pagerange{\pageref{firstpage}--\pageref{lastpage}} \pubyear{2014}

\maketitle

\label{firstpage}

\begin{abstract}
We have observed 99 mid-infrared-bright, massive young stellar objects and compact H{\sc ii} regions drawn from the Red MSX source (RMS) survey in the J=3$-$2 transition of $^{12}$CO and $^{13}$CO, using the James Clerk Maxwell Telescope.  89 targets are within 6\,kpc of the Sun, covering a representative range of luminosities and core masses.  These constitute a relatively unbiased sample of bipolar molecular outflows associated with massive star formation.  Of these, 59, 17 and 13 sources (66, 19 and 15 percent) are found to have outflows, show some evidence of outflow, and have no evidence of outflow, respectively. The time-dependent parameters of the high-velocity molecular flows are calculated using a spatially variable dynamic timescale. The canonical correlations between the outflow parameters and source luminosity are recovered and shown to scale with those of low-mass sources. For coeval star formation we find the scaling is consistent with all the protostars in an embedded cluster providing the outflow force, with massive stars up to $\sim$30\,M$_{\odot}$ generating outflows. Taken at face value, the results support the model of a scaled-up version of the accretion-related outflow-generation mechanism associated with discs and jets in low-mass objects with time-averaged accretion rates of $\sim$10$^{-3}$\,M$_{\odot}$\,yr$^{-1}$ onto the cores. However, we also suggest an alternative model, in which the molecular outflow dynamics are dominated by the entrained mass and are unrelated to the details of the acceleration mechanism. We find no evidence that outflows contribute significantly to the turbulent kinetic energy of the surrounding dense cores.

\end{abstract}


\section{Introduction}
\label{intro}

The formation scenario for massive protostars is comparatively uncertain compared with the low-mass star-formation paradigm \citep{Shu1987}. During their formation, massive stars ($>$\,8\,M$_{\odot}$) deposit copious amounts of energy into the interstellar medium (ISM) and their natal cloud through jets, outflows and expanding ionisation fronts. These processes could act to regulate infall, accretion and local star formation itself.  The well studied massive protostars Cep A HW2 \citep{Patel2005,Curiel2006} and IRAS 20126+4104 \citep{Cesaroni1999,Shepherd2000,Cesaroni2005,Cesaroni2014}, for example, are thought to be surrounded by Keplerian discs and have jets powering massive outflows in a direction perpendicular to the disc plane. The formation scenario for these particular sources appears to be analogous to that of low-mass protostars, although it is unclear whether such sources are representative of all massive protostars, or are isolated cases.

One early signpost of star formation and a potential way to investigate accretion is the phenomenon of bipolar molecular outflows erupting from dense, dark clouds \citep[see][for a more thorough review]{Richer2000,Arce2007}. Early observations summarised in \citet{Lada1985} indicate the vast luminosity range of sources which drive such flows. Subsequent works have confirmed that cores containing massive protostars drive powerful outflows \citep[e.g.][]{Shepherd1996b,Zhang2001, Ridge2001,Beuther2002,Klaassen2007,Mottram2012}.  The outflow phenomenon thereby provides one potential link between low- and high-mass star formation scenarios.

Although detection of outflows alone is insufficient to draw conclusions on the existence of a single star-formation process, correlations between outflow energetics and bolometric luminosity over several orders of magnitude have been interpreted as evidence of a common outflow-driving mechanism that scales with luminosity \citep{Rodriguez1982,Bally1983b,Cabrit1992,Shepherd1996}. 

Few previous studies of outflows from massive protostars have used representative and unbiased samples that are large enough to deal with the observational uncertainties and intrinsic scatter in properties. \citet{Beuther2002} find outflows in 21 of 26 targets and see evidence of similar trends in a sample based on a catalogue of massive star formation sites using IRAS point sources \citep{Molinari1996,Sridharan2002}. Such a sample, containing sources at a range of distances, may be subject to distance biases that overwhelm the significance of the correlation and suffer from confusion in the luminosity values due to the low resolution of IRAS (2-5\,arcmin at 100\,$\mu$m). \citet{Ridge2001} examined a sample of 11 same-distance (2-kpc) sources and found only tentative evidence for a scaling of outflow energetics with bolometric source luminosity in the massive protostar regime, although their sample was small.

\citet{DuarteCabral2013} investigated 9 outflows from mid-IR-dark, high-mass protostellar analogues of Class-0 low-mass sources \footnote{youngest observed YSOs that are deeply embedded (M$_{env}>$M$_{star}$) and have large sub-mm excesses, see \citet{Bontemps1996} and \citet{Lada1987}} at $\sim$1-2 arcsecond resolution and found that the outflow properties scale over $\sim$2.5 orders of magnitude in luminosity. \citet{Sanjosegarcia2013} examined the CO properties of a sample of low, intermediate and high-mass sources, they found a general trend of increasing outflow velocities with bolometric luminosity, but with significant intrinsic scatter. Still, larger samples are required to investigate outflow dynamics and kinematics from representative selection of massive protostellar regions.

The Red MSX Source (RMS) survey \citep{Lumsden2002,Lumsden2013} specifically identifies massive young stellar objects (MYSOs) and H{\sc ii} regions drawn from the MSX mid-infrared survey of the Galactic plane \citep{Egan2003}, which has higher angular resolution (18\,arcseconds) than earlier IRAS point-source catalogues of massive star formation sites \citep{Molinari1996,Sridharan2002} and therefore is less affected by confusion in the Galactic Plane.

This is the second of two papers investigating the same sample of 99 MYSOs and very compact H{\sc ii} regions drawn from the RMS sample. It is the first large sample of molecular outflows from high-mass and high-luminosity YSOs that is not dominated by malmquist-type biases. In many previous attempts to look at high-mass YSOs (e.g. \citealt{Shepherd1996,Beuther2002}, and the high-mass, $\gtrsim$10$^3$\,L$_{\odot}$ component of the definitive \citealt{Cabrit1992} study), accounting for the distance dependencies and possible biases reduces the significance of the correlations.  The conclusion of a single driving mechanism in both low- and high-mass outflow sources therefore requires more careful examination.

In the first of the two papers, \citet{Maud2015} (hereafter Paper I), we identified and analysed the properties of the dense cores in these massive star-forming regions. The luminosities and masses of the cores were shown to be unaffected by significant distance biases out to 6\,kpc, and are representative of the range of luminosities in the whole RMS MYSO/H{\sc ii}-region sample subject to a similar distance cut ($\sim$250 objects). Note we use the terminology `cores' throughout this paper in analogy with other studies of similar high-mass-star formation regions with similar resolutions \citep[see.][]{Beuther2002b,Hill2005}. We probe a range of radii for these dense ($>$10$^4$\,cm$^{-3}$) {\em multiple} star-forming `cores' ($\sim0.1-1.0$\,pc), with a nominal value of $\sim0.35$\,pc (Paper I). As noted in paper I, we understand that there is further sub-structure \citep[e.g.][]{Bontemps2010} which may pertain to {\em single} star-forming cores much smaller than $\sim0.1$\,pc in radius \citep[e.g.][]{Hennemann2009} and all our cores are interpreted as containing young stellar clusters, in which many lower-mass protostars are likely to be co-evolving with the high-mass objects. Furthermore, in terms of a mass-luminosity plot, these regions are indistinguishable from the larger sample; i.e., the MYSOs and H{\sc ii} regions are at a similar evolutionary stage \citep[also see][]{Davies2011,Mottram2011b,Urquhart2014b}. 

Here the JCMT HARP $^{12}$CO and $^{13}$CO (J=3$-$2) observations are specifically used to identify and examine outflows emanating from these regions. Its higher critical density and upper-level energy make the (J=3$-$2) transition an ideal tracer for investigating outflows when compared with the (J=1$-$0) transition, as the entrained outflow emission is warmer than the core and because line-of-sight ambient emission is less prevalent, therefore less likely to contaminate outflow wings. The distance-independent nature and large sample size allows a definitive investigation of outflows from massive protostars. 

Section \ref{obs} summarises the sample and the observations undertaken, Section \ref{meth} describes the outflow identification, parameter determination and outflow parameter calculations. Section \ref{results} presents the general results obtained and Section \ref{full_disc} discusses the observed trends, comparisons with lower-mass outflows, the outflow-driving mechanism and the effect of the outflows on the natal cores. The main results are summarised in Section \ref{summ}.

\section{Sample and Observations}
\label{obs}

Paper I provides a more comprehensive overview of the sample and selection criteria. To summarise here, the 99 sources are all MYSOs and compact H{\sc ii} regions from the RMS sample \citep{Lumsden2013}. The distance limited statistical sample corresponds to 89 of these sources at heliocentric distances less than 6\,kpc. Except where indicated, the luminosities of all sources were calculated using the most up-to-date source distances \citep{Urquhart2012,Urquhart2014} and multi-wavelength spectral energy distribution (SED) fits from \citet{Mottram2011a}.

The observations were undertaken using the James Clerk Maxwell Telescope (JCMT) in 2007 and 2008 as part of projects 07AU08, 07BU16, 08AU19 and 08BU18. The full-width half-maximum (FWHM) beam size at $\sim$345\,GHz for the $^{12}$CO (J=3$-$2) transition is $\sim$14.5\,arcsec. The Heterodyne Array Receiver Program (HARP) 16-pixel SSB SIS receiver \citep{Buckle2009} was used with the ACSIS correlator (Auto-Correlation Spectral Imaging System) backend configured with an operational bandwidth of 1000\,MHz for the $^{12}$CO transition. For the $^{13}$CO data, taken simultaneously with C$^{18}$O (see Paper I), the bandwidth was 250\,MHz. The effective velocity resolutions were $\sim$0.4 ($^{12}$CO) and $\sim$0.06\,km\,s$^{-1}$ ($^{13}$CO and C$^{18}$O). In calculations of the optical depth (see Section \ref{meth}) the $^{13}$CO data is re-sampled to match the $^{12}$CO velocity resolution. Typical spectral noise levels ($\delta T_{\rm mb}$) for both the $^{12}$CO and $^{13}$CO data range between $\sim$\,0.4 and 0.6\,K in a $\sim$\,0.4\,km\,s$^{-1}$ bin. 

Maps of the sources were taken in raster-scan mode with continuous (``on-the-fly") sampling and position switching to observe a `clean' reference position at the end of each scan row. Pointing was checked against a known, bright molecular source prior to each science observation and is likely to be within $\sim$5\,arcsec, typical of the JCMT. Reduction was undertaken with a custom pipeline utilising the \textsc{kappa, smurf, gaia} and \textsc{splat} packages which are part of the \textsc{starlink} software maintained by the Joint Astronomy Centre (JAC)\footnote{http://starlink.eao.hawaii.edu/starlink}. Linear baselines were fitted to the source spectra over emission-free channels and subtracted from the data cubes. Bad baselines on each of the working receivers were flagged if they showed evidence of sinusoidal fluctuations. The final cubes used in the analysis were re-gridded to a 7-arcsec spatial pixel scale. The data, originally on the corrected antenna temperature scale \citep[$T^{*}_{\rm A}$;][]{Kutner1981} were converted to main-beam brightness temperature  $T_{\rm mb} = T^{*}_{\rm A}/\eta_{\rm mb}$, where $\eta_{\rm mb} =$ 0.66 as measured by JAC during the commissioning of HARP \citep{Buckle2009} and via continual planet observations. 

\begin{table*}
\begin{center}
\caption{Source parameters for all objects in the sample, taken from the RMS survey online archive. Only a small portion of the data is provided here, the full table is available in the electronic supplementary information and via the RMS database directly.}
{\footnotesize
\begin{tabular}{@{}llllrrrll@{}}
\hline
MSX Source Name  & RA.   & DEC.   &  Type  &  $\upsilon_{\rm LSR}$  & Distance & Luminosity  & IRAS source  & Other  \\
                & (J2000) & (J2000) &      &   (km\,s$^{-1}$)  &  (kpc)  &  (L$_{\odot}$)  &   (offset)   & Associations \\
\hline
  G010.8411$-$02.5919 & 18:19:12 & $-$20:47:30 & YSO & 11.4 & 1.9 & 24000 & 18162$-$2048 (4$\arcsec$) & GGD27 \\
  G012.0260$-$00.0317 & 18:12:01 & $-$18:31:55 & YSO & 110.6 & 11.1 & 32000 & 18090$-$1832 (3$\arcsec$) & ...\\
  G012.9090$-$00.2607 & 18:14:39 & $-$17:52:02 & YSO & 35.8 & 2.4 & 32000 & 18117$-$1753 (11$\arcsec$) & W33A \\
  G013.6562$-$00.5997 & 18:17:24 & $-$17:22:14 & YSO & 48.0 & 4.1 &14000 & 18144$-$1723 (2$\arcsec$) & ... \\
  G017.6380$+$00.1566 & 18:22:26 & $-$13:30:12 & YSO & 22.5 & 2.2 & 100000 & 18196$-$1331 (11$\arcsec$) & ... \\
  G018.3412$+$01.7681 & 18:17:58 & $-$12:07:24 & YSO & 32.8 & 2.9 & 22000 & 18151$-$1208 (16$\arcsec$) & ... \\
  G020.7438$-$00.0952 & 18:29:17 & $-$10:52:21 & H{\sc ii} & 59.5 & 11.8 & 32000 &...& GRS G020.79$-$00.06 \\
  G020.7491$-$00.0898 & 18:29:16 & $-$10:52:01 & H{\sc ii} & 59.5 & 11.8 & 37000 &...& GRS G020.79$-$00.06 \\
  G020.7617$-$00.0638 & 18:29:12 & $-$10:50:34 & YSO/H{\sc ii} & 57.8 & 11.8 & 62000 &...& GRS G020.79$-$00.06 \\
  G023.3891$+$00.1851 & 18:33:14 & $-$08:23:57 & YSO & 75.4 & 4.5 & 24000 & 18305$-$0826 (6$\arcsec$) & GRS G023.64$+$00.14 \\
\hline
\end{tabular}
}
\label{tab1}
\end{center}
\end{table*}

Table \ref{tab1} lists the sources and their properties as extracted from the RMS online data base \footnote{http://rms.leeds.ac.uk/cgi-bin/public/RMS\_DATABASE.cgi}. In addition, other common source association names are presented along with the IRAS designation and offset if available.

\section{Outflow identification and Parameter determination}
\label{meth}

Outflows are typically identified by characteristic high-velocity wing components in molecular rotational emission-line spectra.  Outflows with axes close to the plane of the sky will not exhibit such broad wings but may have a linear structure in an integrated emission map encompassing near-$\upsilon_{\rm LSR}$ velocities. However, even a known low-inclination, high-mass outflow source in our sample, GGD-27 (HH80-81, G010.8411$-$02.5919), has reasonably broad wings with $\sim$15\,km\,s$^{-1}$ full-width-zero-intensity (FWZI) \citep{Yamashita1989}. 

\begin{figure*}
\begin{center}
\includegraphics[width=0.95\textwidth]{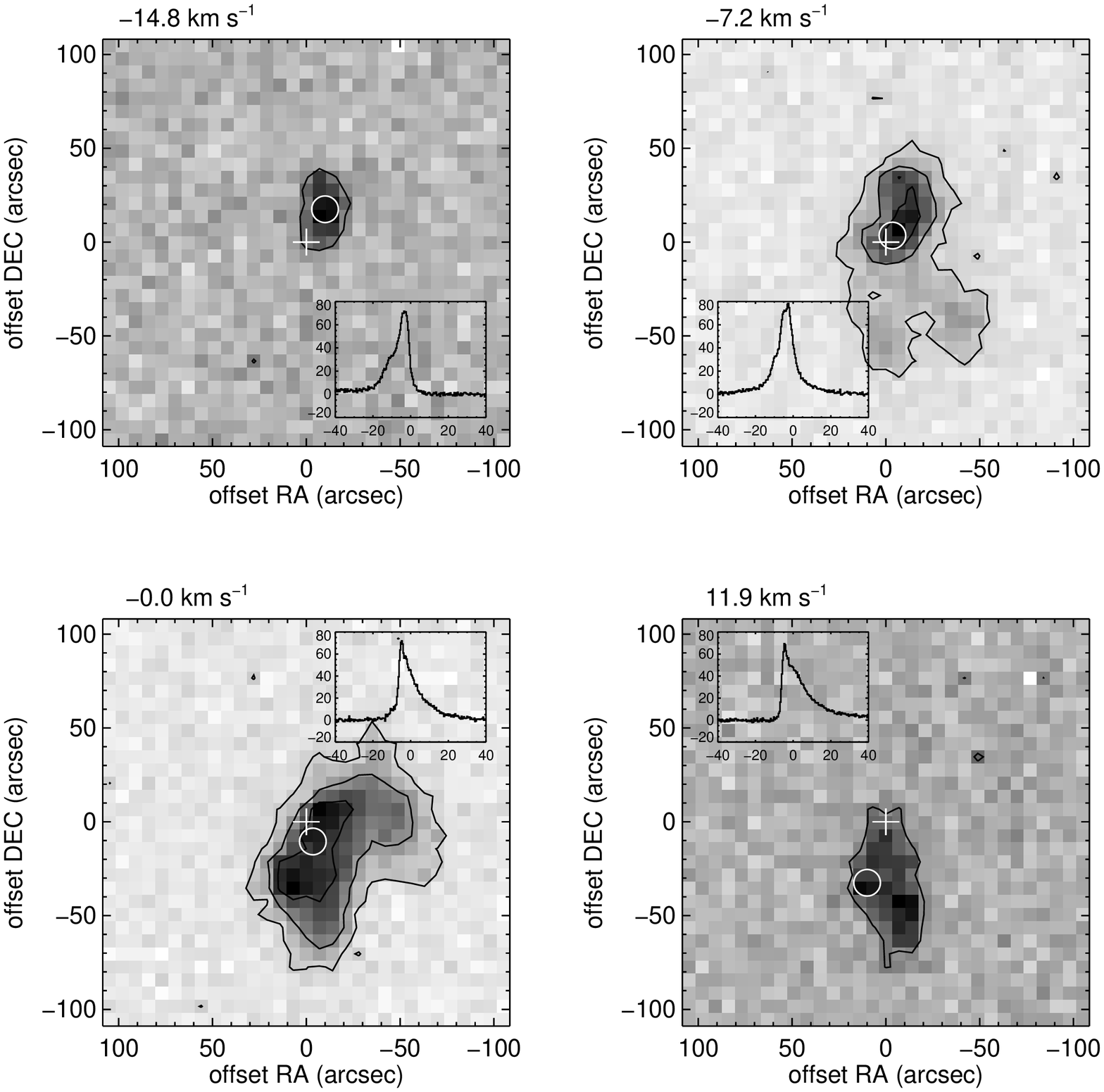}
\caption{Each sub-plot shows a velocity slice (indicated in top left of each sub-plot) from the data cube for G078.1224+03.6320. Inset is the summed spectrum extracted within the white circle. It is clear when moving from blue to red-shifted velocities that there is an outflow with a clear spatial offset about the central source location (white cross).}
\label{fig1} 
\end{center}
\end{figure*}

The data were investigated interactively (using the {\sc gaia} package), by examining slices of the data cube while simultaneously extracting spectra within a 3-pixel diameter circular aperture.  To highlight the outflow emission, a 3\,$\sigma_{T_{\rm mb}}$ cut is applied to the cubes (where $\sigma_{T_{\rm mb}}$ is the standard deviation of the spectral noise measured in emission-free velocity channels extracted at every pixel). Figure \ref{fig1} illustrates the identification process for the source G078.1224+03.6320 (IRAS 20126+4104), a well studied MYSO with a bipolar outflow. Slices of the cube at blue- and red-shifted velocities with respect to the source $\upsilon_{\rm LSR}$ are shown and inset are the spectra extracted from the cube at spatial positions corresponding to the white circle. It is clear that the blue-shifted outflow component lies to the north while the red-shifted material lies to the south \citep[c.f.][]{Shepherd2000}. Note the more chaotic spatial structure closer to the $\upsilon_{\rm LSR}$ velocity ($-$3.6\,km\,s$^{-1}$). This investigation was undertaken for all targets in order to establish the presence or lack of extended velocity material over the core extent found in Paper I as a proxy for an outflow.

\subsection{Temperature Calculation and Optical Depth Correction}
\label{optd}
The properties of the outflows, mass, momentum and energy are determined, essentially, from the moments:

\begin{eqnarray}
\label{eqn0}
\iiint  N(x,y,\upsilon) \upsilon^n \,dx \,dy \,d\upsilon \quad [n=0,1,2]
\end{eqnarray}

\noindent all of which hinge on an accurate determination of the column density $N(x,y,\upsilon)$ (where $x,y$ are spatial coordinate and $\upsilon$ the velocity coordinate in a data cube). To estimate the column density elements $N(x,y,\upsilon)$, the $^{12}$CO (3$-$2) emission is assumed to be in local thermodynamic equilibrium (LTE). Paper I notes that this is likely to be the case for these cores. A single excitation temperature $T_{\rm ex}$ and position- and velocity-dependent optical depth values $\tau$ are calculated for each source. As the $^{13}$CO emission is shown to be optically thick (at the core $\upsilon_{\rm LSR}$) in all but four of these sources (Paper I), it is expected that the $^{12}$CO is also optically thick. This is known to be the case in other outflow sources \citep[e.g.][]{Cabrit1990, Choi1993}.

Thus a single $T_{\rm ex}$ for each source is calculated from:

\begin{eqnarray}
\label{eqn1}
T_{\rm ex} = \frac{16.59 \, {\rm K}}{{\rm ln}[1 + 16.59 \, {\rm K} / \big( {T_{\rm mb,12}} + 0.038 \big)]},
\end{eqnarray}
 
\noindent where ${h\nu}(^{12}{\rm CO})/{k}$ = 16.59\,K, with $\nu(^{12}{\rm CO})$ = 345.80\,GHz and $T_{\rm mb,12}$ is the peak main-beam brightness temperature of the $^{12}$CO emission at the location of the source within one beam and at the source $\upsilon_{\rm LSR}$. Note, in some cases where self absorption occurs the velocity of the peak $T_{\rm mb,12}$ is slightly shifted from the source $\upsilon_{\rm LSR}$.

Often, a constant factor is used to correct outflow masses for optical depth \citep[e.g.][]{Beuther2002}. However, since we have both $^{12}$CO and $^{13}$CO data cubes, we can calculate an optical depth for each voxel and correct every spatial $(x,y)$ and velocity $(\upsilon)$ element by its own optical depth, assuming $T_{\rm ex}$ is the same for both isotopologues, in which case:

\begin{eqnarray}
\label{eqn2}
\frac{T_{\rm mb,12}}{T_{\rm mb,13}} = \frac{1 - e^{-\tau_{12}}}{1 - e^{-\tau_{13}}} = \frac{1 - e^{-\tau_{12}}}{1 - e^{(-\tau_{12}/{R})}},
\end{eqnarray}

\noindent where $R$ is the abundance ratio of [$^{12}$CO]/[$^{13}$CO] = $7.5 \times {D_{\rm gc}}+7.6$, following \citet{Wilson1994}, and $D_{\rm gc}$ is Galactocentric distance in kpc. Equation \ref{eqn2} can be used effectively where the $^{13}$CO emission is optically thin, and although $\bar{\tau}_{\rm 13, core} \sim 3.5$ (Paper I), $\tau_{13} < $1 in the line wings away from the source $\upsilon_{\rm LSR}$.

\citet{Maud2013T} found that the mean of all the individual voxel $(x,y,\upsilon)$ optical depths, $\bar{\tau}_{12}$, is consistent with that derived using $\sum(\int T_{\rm mb,12} \, d\upsilon)$ and $\sum(\int T_{\rm mb,13}\, d\upsilon)$ directly in place of $T_{\rm mb,12}$ and $T_{\rm mb,13}$ in Equation \ref{eqn2}. However, later calculations of momentum and energy, for example, can become grossly overestimated (by orders of magnitude) using such a single, averaged optical-depth correction, because the higher velocity, optically thin emission is over-corrected.

\subsection{Outflow Masses}
\label{optmass}

Once the optical depth and excitation temperature are established for each voxel, the column density maps for each outflow lobe, $N(x,y)$ pixels, are calculated via: 

\begin{eqnarray}
\label{eqn3}
N({\rm ^{12}CO}_{\rm x,y}) = 4.78\, {\times}\, 10^{12} \, \frac{{\rm exp}(16.74/T_{\rm ex}) \, (T_{\rm ex} + 0.93)}{{\rm exp}(-16.59/T_{\rm ex})} \, \nonumber\\  \times \, \int T_{\rm mb,12} \, \frac{\tau_{12}}{[1 - {\rm exp} (-\tau_{12})]}\, d\upsilon\;,
\end{eqnarray}

\noindent where ${N}({\rm C^{12}O}) $ is in ${\rm cm^{-2}}$ and $d\upsilon$ is in this case is the integration over the velocity extent in \,km\,s$^{-1}$; hence, the mass elements at each pixel coordinate, $x,y$ (where $n$ = 0 in Equation \ref{eqn0}) are,

\begin{eqnarray}
\label{eqn4}
M_{\rm gas, ({\rm x,y})} = {N}({\rm ^{12}CO}_{\rm x,y}) \bigg[ \frac{{\rm H_2}}{\rm ^{12}CO} \bigg] \mu_{g}\,m(_{\rm H_2})\Omega\,D^2
\end{eqnarray}

\noindent where $\Omega$ is the solid angle of a pixel element (x,y), $D$ is the distance to the source, (H$_2$/$^{12}$CO) is the H$_2$ to $^{12}$CO abundance ratio = 10$^{4}$ and $\mu_{g}$\,=\,1.36 is the total gas mass relative to H$_2$. A more detailed derivation of the column density and mass are given in Appendix \ref{AppendixA}.

Calculation of the outflow mass, which then leads to momentum and energy (where $n$ = 1,2 in Equation \ref{eqn0}), however, first relies on the correct identification of emission associated with the high-velocity gas in order to reliably separate it from that of the core. The method used here is analogous to that outlined in Paper I, a combined process of integration over velocity (i.e. creation of zeroth-order moment maps for each outflow lobe) followed by an aperture summation over the area of outflow emission. The blue and red-shifted outflow velocity extent (maximal velocities) are established via a direct, manual, investigation of the raw data cubes, channel by channel, until emission spatially associated with the outflow drops below 3$\sigma_{T_{\rm mb}}$ within a beam area. Automating this process for a large sample is difficult as there is typically diffuse emission at `outflow' velocities elsewhere in the maps that is not associated with the outflow itself.

\begin{table*}
\begin{center}
\caption{Outflow-detection parameters for all objects in the sample. Y, M, N in the flow column indicates whether the source has an outflow, shows some evidence of an outflow (maybe), or no outflow at all. Y, S, N in the Aperture Flag column indicates a good aperture, a manually `selected' aperture or no aperture. Sources without outflows have no aperture, but also some M sources have no aperture due to complex diffuse emission. $\Delta \upsilon$ is the raw velocity extent of the lobes with respect to the observed velocities, whereas $\upsilon_{\rm max}$ are the maximum velocity offsets from the $\upsilon_{\rm LSR}$. The full version of this table is available in the electronic supplementary information.}
{\scriptsize
\renewcommand{\tabcolsep}{1.5mm}
\begin{tabular}{@{}lccrrrrrrrrc@{}}
\hline
MSX Source Name  & Flow   & Spec.  &  $\Delta\upsilon_{\rm blue}$ &  $\Delta\upsilon_{\rm red}$  & Blue Map & Blue Map & Red Map  & Red Map  &  $\upsilon_{\rm max, b}$ &$\upsilon_{\rm max, r}$  & Aper.    \\       
                 &     Flag       & Noise     &                              &    & Med. Noise & ($\sigma$) & Med. Noise & ($\sigma$) &  &  & Flag  \\
                &            & (K)       & (km\,s$^{-1}$) & (km\,s$^{-1}$) &(K\,km\,s$^{-1}$) &(K\,km\,s$^{-1}$)&(K\,km\,s$^{-1}$)&(K\,km\,s$^{-1}$) & (km\,s$^{-1}$)& (km\,s$^{-1}$) & \\
\hline
  G010.8411$-$02.5919   &  Y   &  0.4   &  ( $-$6.1  , 10.0   )   &   ( 14.2  , 28.2   )   &  0.1  &  1.2  &  3.4  &  1.1  & 18.2  & 16.0 &  Y    \\
  G012.0260$-$00.0317   &  Y   &  0.9   &  ( 93.9  ,108.1   )   &   (113.8  ,123.1   )   &  6.7  &  2.2  &  1.0  &  1.8  & 17.0  & 12.2 &  Y    \\
  G012.9090$-$00.2607   &  Y   &  0.5   &  (  7.0  , 32.3   )   &   ( 40.4  , 73.0   )   & 33.2  &  1.7  & 44.1  &  1.9  & 29.4  & 36.7 &  S    \\
  G013.6562$-$00.5997   &  Y   &  0.8   &  ( 28.4  , 45.0   )   &   ( 50.6  , 63.1   )   & 15.9  &  2.0  &  6.6  &  1.8  & 19.4  & 15.3 &  Y    \\
  G017.6380$+$00.1566   &  M   &  0.5   &  (  6.0  , 20.0   )   &   ( 25.9  , 38.2   )   & 10.8  &  1.2  & 11.1  &  1.1  & 16.3  & 15.9 &  S    \\
  G018.3412$+$01.7681   &  Y   &  0.5   &  ( 19.0  , 31.3   )   &   ( 35.9  , 50.3   )   &  2.2  &  1.0  &  1.3  &  1.1  & 13.8  & 17.6 &  S    \\
  G020.7438$-$00.0952   &  M   &  0.5   &  ( 46.9  , 56.1   )   &   ( 62.1  , 69.3   )   & 10.4  &  1.1  &  7.0  &  0.9  & 12.2  & 10.2 &  S    \\
  G020.7491$-$00.0898   &  M   &  0.5   &  ( 46.9  , 55.4   )   &   ( 62.3  , 69.3   )   &  5.4  &  1.0  &  7.0  &  0.9  & 11.9  & 10.5 &  S    \\
  G020.7617$-$00.0638   &  Y   &  0.5   &  ( 43.9  , 53.1   )   &   ( 59.6  , 82.0   )   &  0.6  &  1.1  & 35.0  &  1.6  & 12.5  & 25.6 &  Y    \\
  G023.3891$+$00.1851   &  Y   &  0.5   &  ( 58.2  , 73.5   )   &   ( 77.2  , 91.2   )   &  3.8  &  1.2  & 22.6  &  1.1  & 17.2  & 15.9 &  Y    \\
       \hline
\end{tabular}
\renewcommand{\tabcolsep}{2.1mm}
}
\label{tab2}
\end{center}
\end{table*}

\citet{Ridge2001} note that careful removal of the core component is required to determine outflow properties accurately. The low-velocity limits of each outflow are established with reference to the velocity extent of the core component in Paper I. We use the largest of: (1) the established velocity range (Paper I, Table 3); or (2) the full-width-tenth of maximum (FWTM = 1.8 $\times$ C$^{18}$O FWHM) as this ensures that most of the core emission is excluded from that associated with the outflow lobes. Generally, case 1 $\sim$ case 2 wherever the C$^{18}$O emission is well above 3\,$\sigma_{T_{\rm mb}}$.  However, this is not the case for all the cores due to varying noise levels; hence the rationale of using the maximum of case 1 or 2. Furthermore, the difference between case 1 and case 2 is not more than 1.0\,km\,s$^{-1}$, which is accounted for in the uncertainty analysis. At most, we find the mass of the outflow could be underestimated by 50 percent.  However, the influence on the momentum and energy parameters is smaller as the `missed' emission is at very low velocities and contributes much less to values containing multiples of the velocity. The velocity ranges established by this process are listed in Table \ref{tab2}, including the spectral noise, $\sigma_{T_{\rm mb}}$, and the maximum velocities shifted by the source $\upsilon_{\rm LSR}$.

After integration over the selected velocity range, a polygon aperture is used to sum the spatial extent of the outflow emission in the integrated lobe maps. A problem with this method is that true plane-of-the-sky outflows (i.e. with no wing components) will be excluded because their velocity range is wholly within that attributed to the core. However, we do not identify any plane-of-sky outflows in the cube investigation stages, where we search the data cubes directly for evidence of elongated features close to the core $\upsilon_{\rm LSR}$.

The polygon aperture used to define the spatial extent of an outflow is adapted from the simple 3\,$\sigma_{\rm MAP}$ contour method established in Paper I. The $\sigma_{\rm MAP}$ is calculated from the average spectral noise, and is equivalent to a spatial average of the physical map noise (see Appendix B), but we find that only in a handful of cores, with little to no ambient $^{12}$CO emission in the regions surrounding the outflow lobes, does 3\,$\sigma_{\rm MAP}$ correctly delineate the outflow lobe region (e.g. as in G078.1224$+$03.6320). In more complex regions, with ambient emission away from the main outflow, we can use the median value of the map to act as a background level (note the median map values are $\sim$\,zero, for sources with no ambient emission, such as G078.1224$+$03.6320), and define the aperture delineating the outflow lobes to follow the contour of 3\,$\sigma_{\rm MAP}$ above this background (see Appendix B for details).

\begin{figure*}
\begin{center}
\includegraphics[width=0.95\textwidth]{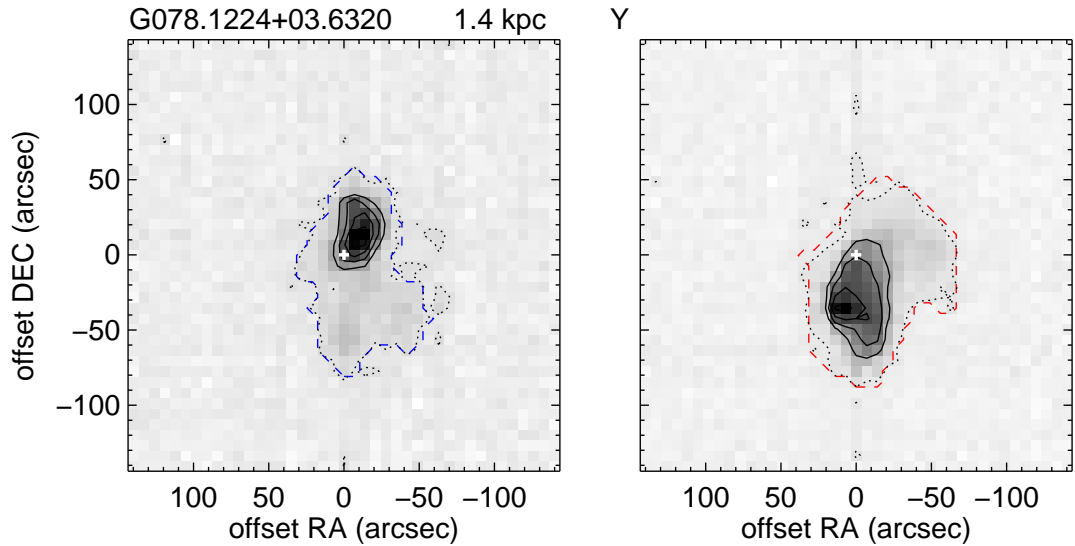}
\caption{Left: Map of integrated intensity  in the blue-shifted outflow lobe of G078.1224+03.6320, prior to optical depth correction. The integration velocity range is $-$40.0 to $-$5.7\,km\,s$^{-1}$. The blue dashed line indicates the aperture at the contour equivalent to 3\,$\sigma_{\rm MAP}$ above the background (defined by the median map value). The black contours indicate the 30, 50, 70 and 90 percent levels to highlight offset of the emission from the source location indicated by the white cross. Right: As left but for the red-shifted outflow lobe in the range is $-$1.0 to 40\,km\,s$^{-1}$. The source $\upsilon_{\rm LSR}$ is $-$3.6\,km\,s$^{-1}$.  The source name, distance and outflow flag are indicated at the top of the plot, Y, in this case represents a good outflow source with well defined apertures.}
\label{fig2} 
\end{center}
\end{figure*}

However, we note that, in some sources the outflow areas are not well defined by this method and the apertures do have to be manually adjusted to divide the diffuse emission from that of the outflow. These sources are flagged as `selected' (S) apertures in Table \ref{tab2}. Appendix C presents the outflow lobe maps for each source and the apertures used for area summation, while Figure \ref{fig2} shows the example of G078.1224+03.6320. Note that apertures are set prior to optical depth correction as they are based on the observational noise levels. All the appendices are available in the additional online material.

\subsection{Outflow Parameters}
\label{params}

Empirical tests conducted as part of the analysis investigated a number of methods used in the literature to establish the outflow parameters \citep[see][for more details]{Maud2013T}. The most accurate in calculating the momentum, $P$, and energy, $E$, are:

\begin{eqnarray}
\label{eqn5}
P = \sum_{x,y,i} \,M_i v_i
\end{eqnarray}

\begin{eqnarray}
\label{eqn6}
E = \frac{1}{2}\sum_{x,y,i} \,M_i v_i^2
\end{eqnarray}

\noindent where $M$ is the mass, $i$ represents each velocity bin, at velocity $\upsilon_i$ with respect to the source $\upsilon_{\rm LSR}$ summed over all velocities and spatial pixels for both outflow lobes (as Equation \ref{eqn0} with $n$ = 1 and 2). These are the most accurate in recovering input parameters when tested on outflow models \citep{Cabrit1990}. This method of \emph{full cube analysis} ($x,y,i$) avoids the overestimate of parameters that occurs when the total outflow mass is simply multiplied by the maximum velocity \citep[e.g.][]{Beuther2002,Lebron2006}, a method that notionally places all the mass at the maximal outflow velocity (\citealt{Margulis1985}).

\begin{figure*}
\begin{center}
\includegraphics[width=0.80\textwidth]{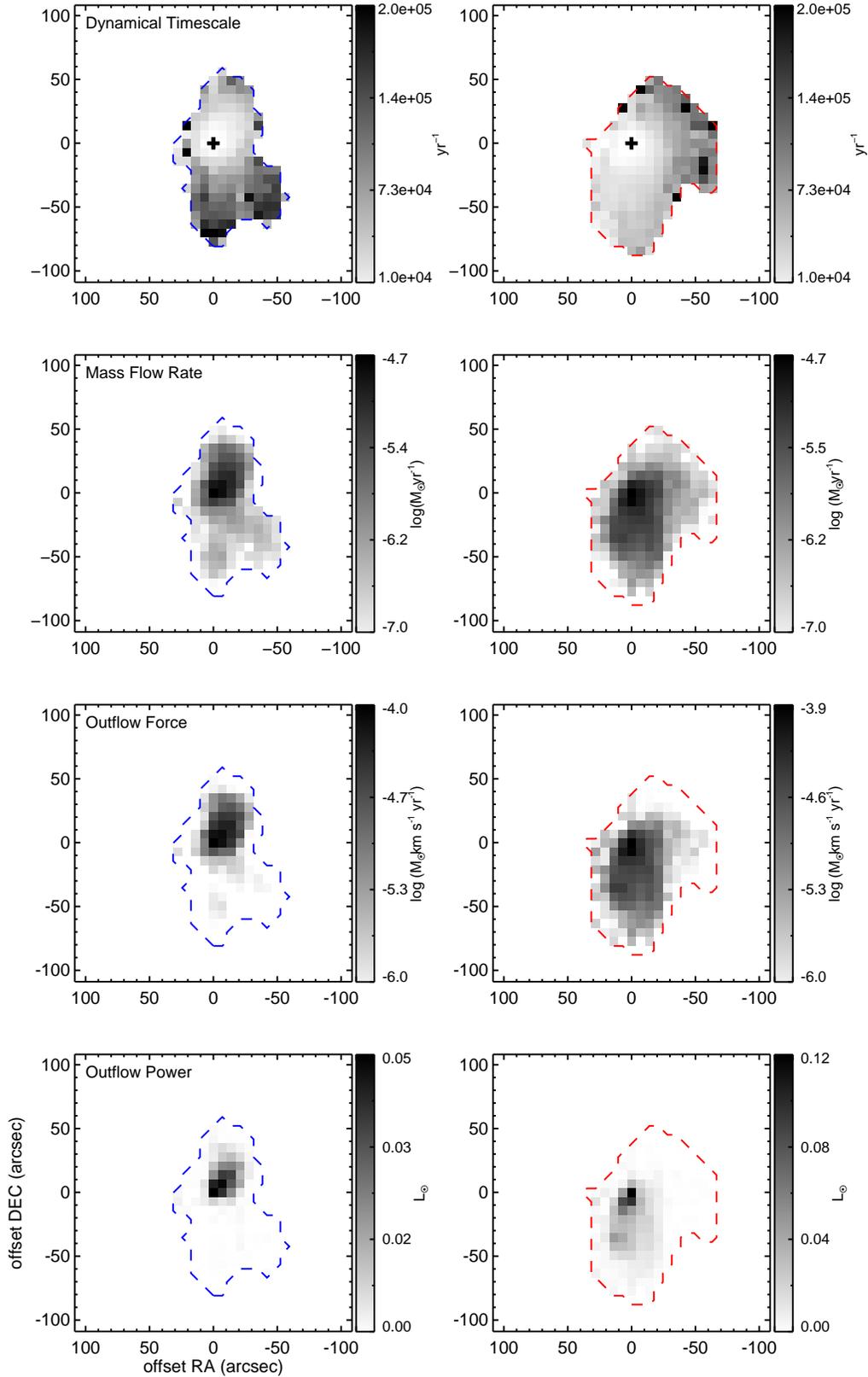}
\caption{Top$-$Bottom: The spatial distributions of the calculated dynamical timescale, mass flow rate, force and outflow power (luminosity). The left and right panels show the blue- and red-shifted outflow lobes, respectively. With reference to Figure \ref{fig2} the positionally offset, higher-velocity emission corresponds to lower dynamical timescales. The mass flow rate, force and power are strongest in regions of low $t_{{\rm dyn}(x,y)}$, as expected. Lower velocity diffuse material in the case of G078.1224+03.6320 contributes little to the flow, force and power as these have the largest $t_{{\rm dyn}(x,y)}$. Note the logarithmic scaling for flow and force to show the weak emission.}
\label{fig3} 
\end{center}
\end{figure*}

In order to establish the mass flow rate, $\dot{M}_{\rm out}$, the force (momentum supply rate), $\dot P$ = $F_{\rm m}$, and the mechanical luminosity (outflow power), $\dot E$ = $L_{\rm m}$, a dynamical timescale $t_{\rm dyn} = R/\upsilon$ must be estimated, where $R$ is the distance of the outflow from the source. The dynamic timescale is commonly taken to be an indicator of the age of the outflow and, as such, could also be representative of the protostar's age \citep[see,][]{Beuther2002} (under the assumption of a constant outflow over time that begins as soon as the protostar is formed). Further interpretation is discussed below in Section \ref{tdyn}. \citet{Lada1996} introduced a means to calculate $t_{\rm dyn}$ at all spatial positions using $R_{x,y}/\langle \upsilon_{x,y} \rangle$, where $\langle V_{x,y} \rangle$ is the intensity-weighted-mean outflow lobe velocity, representative of the bulk motion of material, and calculated from $\langle P_{x,y}/M_{x,y} \rangle$. To utilise our outflow lobe maps we adopt this method. From the position-dependent $t_{\rm dyn}$, the dynamic parameters are calculated via $\dot{M}_{{\rm out}(x,y)} = M_{(x,y)}/t_{{\rm dyn}(x,y)} $, $\dot P_{(x,y)} = F_{{\rm m}(x,y)} = P_{(x,y)}/t_{{\rm dyn}(x,y)} $ and $\dot E_{(x,y)} = L_{{\rm m}(x,y)} = E_{(x,y)}/t_{{\rm dyn}(x,y)} $. In Figure \ref{fig3} it is clear that the largest contributor to $\dot M$, $\dot P$ and $\dot{E}$ are the spatially offset, higher-velocity outflow components. The spatially diffuse, chaotic, low-velocity emission contributes very little ($\sim$ a few percent) to the total values.

From all outflow parameters the mass is the only one independent of velocity. The velocity dependent variables (and those involving outflow length), in principle, should be corrected for the source inclination angle as we are only sensitive to the line-of-sight velocity component (or that measured on the sky, in terms of lengths). In light of this: the momentum and energy will be $\sim$lower limits (both $\propto$\,1/cos$^n\,\theta$, where $n$\,=\,1 and 2 respectively, and where $\theta$ is the inclination angle with respect to the line of sight); the dynamical timescale will be over- or under-estimated if $\theta$ is $>$\,45$^{\circ}$ or $<$\,45$^{\circ}$ respectively ($\propto$\,1/tan$\,\theta$); $\dot{M}_{\rm out}$ follows the inverse relationship of $t_{\rm dyn}$; and $F_{\rm m}$ and $L_{\rm m}$ will be over- or under-estimated where $\theta$ is less-than or greater-than $\sim$38$^{\circ}$ ($\propto$\,sin\,$\theta$/cos$^n\,\theta$ where $n$\,=\,2 and 3). Inclination angles are not available for these sources and cannot easily be established. \citet{vandermarel2013} apply inclination angle corrections for their sample of low-mass outflows, using the notion of pole-on, plane-of-sky, and somewhere between; often however, a constant correction factor is used for a nominal inclination of $\sim$57.3$^{\circ}$ \citep[e.g][]{Bontemps1996}. We do not apply any factors unless specifically noted (see Sections \ref{mme} and \ref{lowM}).

\section{Results}
\label{results}

\begin{figure*}
\begin{center}
\includegraphics[width=0.95\textwidth]{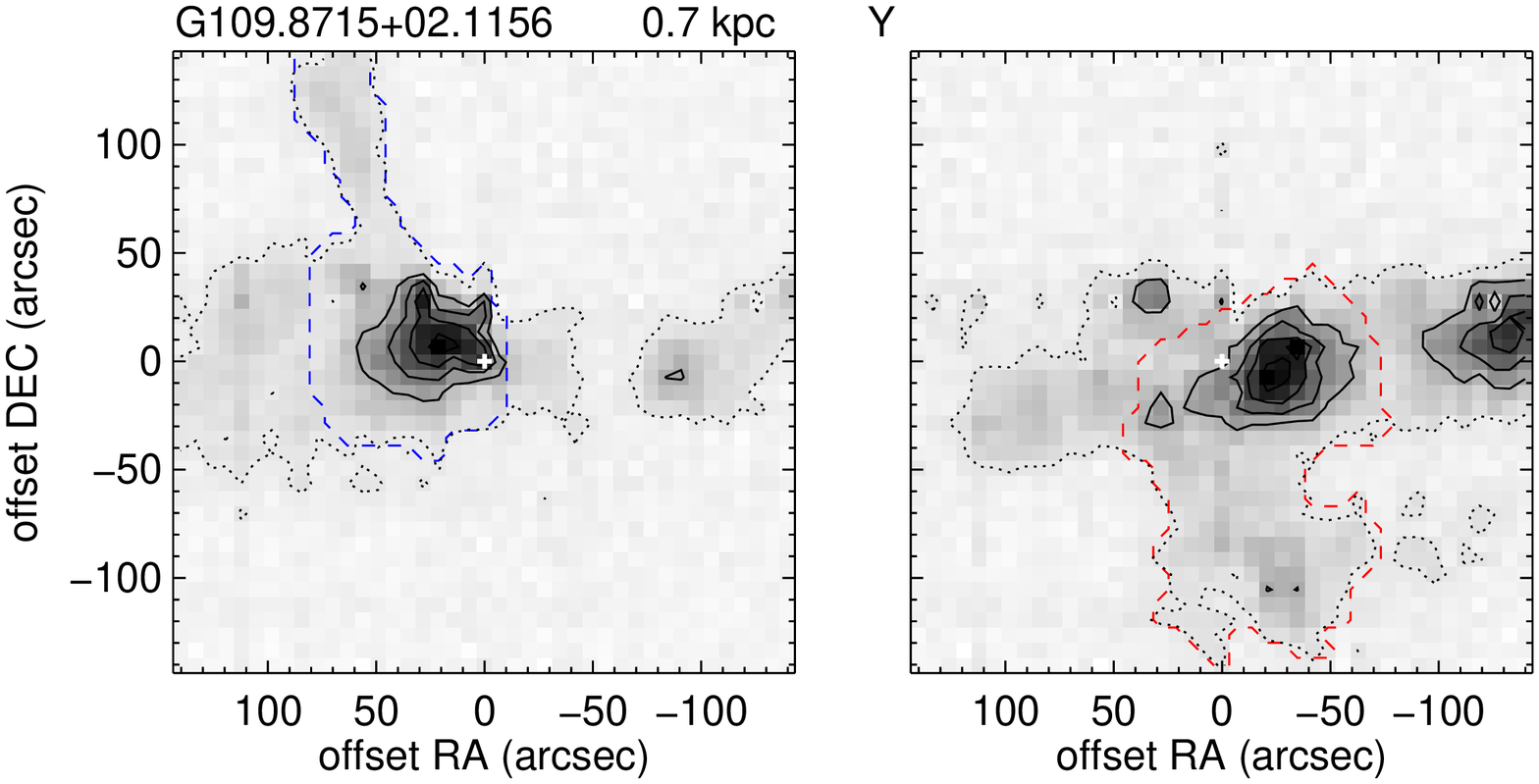}
\caption{As Figure \ref{fig2} but for the source G109.8715$+$02.1156. The velocity ranges are listed in Table \ref{tab2}. This source contains at least two outflows, one close to the plane of the sky (extension in a roughly North-South direction), and another, with broader outflow wings, in an East-West direction. Given the spatial resolution, separation of these components is subjective and hence the total `cluster' outflow parameters are calculated (see Section \ref{lowM}).}
\label{figCepA} 
\end{center}
\end{figure*}

From the total sample of 99 sources 65 are associated with molecular outflows, all of which exhibit line-wings in their $^{12}$CO spectra. 14 show no evidence of outflow identifiers such as high-velocity wings or plane-of-the-sky linear structures - their $^{12}$CO and $^{13}$CO spectra are purely Gaussian in profile. 20 sources in complex regions (spatially and spectrally) show some high-velocity components but also multiple spectral peaks which confuses the outflow identification. In the distance-limited subset (89 sources) there are 59 definite outflows, 17 with outflow like properties and 13 with no outflow evidence. Not all of the sources with outflows show a clear spatial offset of the blue and red-shifted velocity components like G078.1224+03.6320. Given the resolution of the single-dish observations, it is not expected that all spatially offset velocity components would be resolved. There is evidence, however, for a distribution of source inclinations, as some of the most distant sources, expected to be the least resolved, do exhibit clear spatially offset blue- and red-shifted outflow lobes (e.g. G053.5343$-$00.7943 at 5\,kpc). Only two sources appear to have clear outflows close to the plane of the sky associated with the cores (the aforementioned G010.8411$-$02.5919 and G109.8715$+$02.1156, although these cores drive multiple distinguishable outflows, see Figure \ref{figCepA}). This can also be confirmed as the sources have been previously well studied at radio wavelengths and have very linear radio jets (GGD 27, \citealt{Marti1993} and Cep A HW2, \citealt{Curiel2006}). G203.3166$+$02.0564 also has a close to plane-of-sky outflow forming a linear structure across the map. This outflow is offset from the source location to the East and the velocity range is optimised only for the outflow associated with our source position (also see \citealt{Maury2009}, Cunningham et al. 2015, submitted to MNRAS). However, as previously noted we do not identify any isolated, true plane-of-the-sky outflows. It is inherently difficult in these sources to spatially separate the complex core emission from potential plane-of-the-sky outflows and so it is possible that some may be missed as they are simply confused by the core emission.

Table \ref{tab3} lists the masses, momenta and energies while Table \ref{tab4} lists the dynamical-time-dependent parameters, mass flow rate, force and power (luminosity). The reported values of $\dot M_{out}$, $\dot P$ and $\dot E$ are those summed within the defined apertures (blue and red dashed contours in Figures \ref{fig2} and \ref{fig3}), although the $t_{\rm dyn}$ values in Table \ref{tab4} are those calculated directly from $R_{\rm max}/\upsilon_{\rm max}$, as one cannot establish a meaningful single value from a spatially variable map (note the values obtained for $\dot M_{\rm out}$, $\dot P$ and $\dot E$ using this single $t_{\rm dyn}$ are closely comparable with those using presented in Table \ref{tab4} using the spatially variable $t_{{\rm dyn}(x,y)}$, see Section \ref{lowM}).

Figure \ref{fig5} shows the dependence of the calculated outflow parameters $M$, $P$, $E$, $\dot M$, $\dot P$ and $\dot E$ on the bolometric luminosity of the associated MYSO or H{\sc ii} region in the RMS catalogue (shown in Table \ref{tab1}). These plots also include the few sources where D$>$6\,kpc, although they are not used in the statistics or analysis. Table \ref{tab5} provides a list of all Spearman rank correlation statistics for relationships in Figure \ref{fig5}, using only the sources in the distance-limited subsample. Table \ref{tab6} gives the parameters of the corresponding linear fits to the data, again using only the distance-limited sources.

\section{Discussion}
\label{full_disc}

\subsection{Mass, Momentum and Energy}
\label{mme}

\begin{figure*}
\begin{center}
\includegraphics[width=0.82\textwidth]{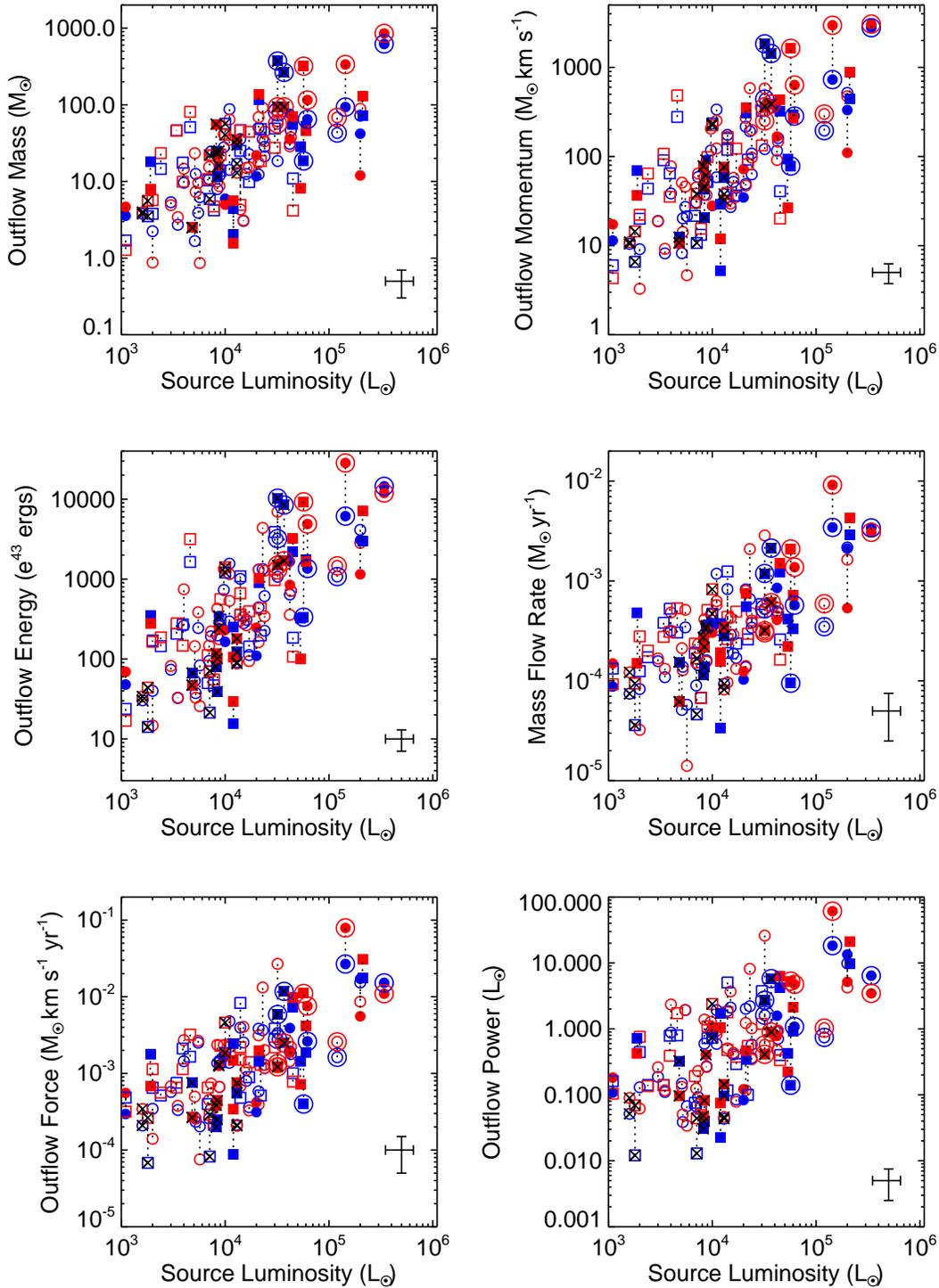} 
\caption{Top$-$Left to Bottom$-$Right, plots of outflow mass, momentum, energy, mass flow rate, force and power against the source luminosity. The blue and red symbols indicate the blue- and red-shifted outflow lobe values and are joined by a dashed line for each source. Open and filled symbols are MYSOs and H{\sc ii} regions, respectively, while circles and squares are for sources with `Y' and `S' aperture flags. The sources that have uncertain outflow evidence and flagged as `M' are indicated by the $\times$ symbols. The source symbols that are circled are those with D$>$6\,kpc. The correlation of all parameters with source luminosity is clear.  Errors for mass, momentum and energy those calculated by increasing and decreasing the integration velocity range by one bin ($\sim$0.4\,km\,s$^{-1}$) at the upper and lower limit, while those indicated for flow, force and power are 50 percent error bars. These uncertainties are direct from the change in integrated velocity range and do not account for uncertainties in source distance, or more importantly the outflow inclination (except mass), and should be considered as minimal uncertainties. As noted in the text, \citet{Cabrit1990} indicate that uncertainties in the energy can be up to an order of magnitude for large inclination angles ($>$\,70$^{\circ}$). The error shown for source luminosity is 30 percent \citep{Mottram2011a}.}
\label{fig5} 
\end{center}
\end{figure*}

The outflow mass is the only velocity-independent variable and so the only one not affected by source inclination angle. The top-left panel of Figure \ref{fig5} indicates that the cores harbouring the most luminous protostars are those with the most massive outflows.  Both the blue- and red-shifted outflow lobe masses are plotted to show that both follow the same trend, and to investigate any possible systematic trend or differences between the lobes. We continue to plot both lobe properties thought the analysis. Even though asymmetric or even single-lobed outflows are not uncommon, in most cases where both blue- and red-shifted lobes are present, the masses are consistent within a factor of $\sim$2$-$3. This is consistent with what one would expect if outflows are entrained material in an isotropic core, for example. A few sources have very asymmetric profiles and, in two particular cases (G023.7097+00.1701 and G050.2213$-$00.6063), there is an absence of high-velocity emission in the blue and red lobes, respectively. There is no clear preference for more massive blue- or red-shifted outflows in the sample. As discussed in Paper I, these cores are clusters of protostars, each of which could be powering outflows that combine and interact such that we observe a preferentially stronger blue- or red-shifted flow. Furthermore, dependent on the core geometry or density distributions the material constituting to the outflow itself may vary considerably (also see Section \ref{tdyn}).

\begin{table*}
\begin{center}
\caption{Mass, momentum and energy calculated for all sources (including those where D$>$6\,kpc) with outflows or with evidence of outflows where apertures could be defined. The `b', `r' and `total' subscripts indicate the blue-shifted lobe, red-shifted lobe and total values of each parameter. The mass is given in units of M$_{\odot}$, momentum in M$_{\odot}$\,km\,s$^{-1}$ and energy in 10$^{43}$\,erg. Uncertainties calculated from adoption a different integrated velocity range ($\pm$\,$\sim$0.4\,km\,$^{-1}$ at the upper and the lower velocity boundary) are $\sim$36, 26 and 23 percent for mass, momentum and energy respectively. Note, errors in source distance, and more importantly outflow inclination angle could have a much more significant effect. \citet{Cabrit1990} suggest uncertainties up to an order of magnitude for energy values if the outflows have large inclination angles ($>$\,70$^{\circ}$). The full table is available online.} 
{\footnotesize
\begin{tabular}{@{}lrrrrrrrrrrrr@{}}
\hline
MSX Source Name && $M_{b}$ & $M_{r}$ & $M_{total}$ && $P_{b}$ & $P_{r}$ & $P_{total}$ && $E_{b}$ & $E_{r}$ & $E_{total}$ \\
\hline 
  G010.8411$-$02.5919  &&   21.3  &   33.4  &   54.7  &&    62.9  &    97.1  &   160.0  &&   225.2  &   342.1  &   567.4   \\
  G012.0260$-$00.0317  &&   81.0  &   56.4  &  137.4  &&   440.8  &   251.0  &   691.8  &&  3201.9  &  1306.8  &  4508.8   \\
  G012.9090$-$00.2607  &&   18.5  &   60.9  &   79.3  &&   121.1  &   580.2  &   701.3  &&  1068.8  &  6948.0  &  8016.9   \\
  G013.6562$-$00.5997  &&   35.8  &    4.3  &   40.1  &&   174.9  &    32.0  &   206.9  &&  1005.1  &   301.6  &  1306.6   \\
  G017.6380$+$00.1566  &&   56.6  &   40.7  &   97.3  &&   235.6  &   224.7  &   460.3  &&  1209.6  &  1409.0  &  2618.6   \\
  G018.3412$+$01.7681  &&   48.3  &   18.5  &   66.8  &&    92.4  &    81.8  &   174.2  &&   197.2  &   430.8  &   628.0   \\
  G020.7438$-$00.0952  &&  372.3  &   94.1  &  466.4  &&  1837.5  &   361.8  &  2199.3  && 10296.9  &  1490.0  & 11786.9   \\
  G020.7491$-$00.0898  &&  264.7  &   94.1  &  358.8  &&  1429.9  &   388.2  &  1818.1  &&  8438.5  &  1699.7  & 10138.2   \\
  G020.7617$-$00.0638  &&   64.7  &  115.6  &  180.2  &&   284.7  &   634.2  &   919.0  &&  1359.2  &  4884.6  &  6243.8   \\
  G023.3891$+$00.1851  &&   28.4  &   44.7  &   73.0  &&   108.0  &   198.7  &   306.8  &&   612.8  &  1276.9  &  1889.7   \\

 \hline
\end{tabular}
}
\label{tab3}
\end{center}
\end{table*}

The next two panels (top right and left middle) of Figure \ref{fig5} show the same clear power law trends of increasing momentum and energy in more luminous protostars seen in many previous studies. The simplest interpretation is that the jet or wind from the most luminous protostar in each core is able to entrain more of the available core material and thus drive the most powerful and most energetic outflow in the region.  Assuming a young stellar cluster is present in these sources, and that its luminosity is dominated by the most massive cluster member, then these relationships support the idea of an outflow-driving mechanism that scales up to protostars as massive as $\sim$50\,M$_{\odot}$ ($\sim$5$\times$10$^{5}$\,L$_{\odot}$), our most massive source (including those more distant than 6\,kpc).

There is no apparent difference in mass, momentum or energy derived for sources classed in the RMS survey as YSO or H{\sc ii} regions, even though these are often thought of as different evolutionary states. \citet{Urquhart2014b} find no difference in the $L_{\rm bol}$ versus $M_{\rm core}$ distribution for these two classifications. We also noted in Paper I that the core properties of these sources are indistinguishable and so they are likely to be at roughly the same evolutionary stage, with similar outflow properties.

Some of the additional scatter in the momentum and energy values in Figure \ref{fig5} is likely caused by a distribution of outflow inclination angles.
The spatial resolution of these single-dish observations does not allow inclination angles to be established. The mean inclination angle of $\sim$57.3$^{\circ}$ in \citet{Bontemps1996} would results in only a constant scaling of $\sim$1.85 (1/cos$\, \theta$) for momentum and $\sim$3.43 (1/cos$^2\, \theta$) for energy, and therefore does not change the relative relationships seen in Figure \ref{fig5} i.e. in log-log plots. \citet{Cabrit1990} however, discuss how outflows with large inclination angles, $>$70$^{\circ}$, could have energies underestimated by an order of magnitude. Furthermore, as already distinguished in two of our cores, there are likely multiple outflows driven by the sources within the cores, and hence a single inclination angle may not be fully representative.

\subsection{Dynamical Timescales}
\label{tdyn}
\begin{figure}
\begin{center}
\includegraphics[width=0.45\textwidth]{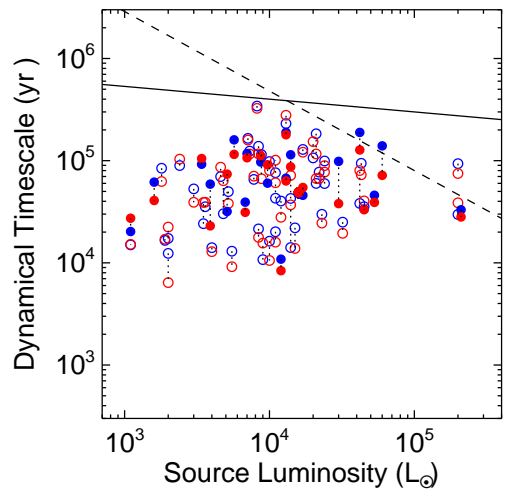} 
\caption{The dynamical timescales calculated using the single-value method, $R_{\rm max}/\upsilon_{\rm max}$, versus the source luminosity. The blue- and red-shifted timescales are $\sim$constant with luminosity, there is no significant correlation between the two parameters. The open and filled symbols represent YSOs and H{\sc ii} regions. Only sources with D$<$6\,kpc are plotted. The dashed and solid lines are the MYSO and H{\sc ii} region phase lifetimes from \citet{Mottram2011b}. }
\label{fig6} 
\end{center}
\end{figure}

The use of $t_{\rm dyn}$ assumes that the local gas velocity is equal to the velocity of the shock wave driven through the molecular gas by the underlying wind or jet, which is the case for an isothermal shock.  Since the shock velocity is likely to be position-dependent, the \citet{Lada1996} position-variable $t_{\rm dyn}$ method is more physical but does not easily allow a single characteristic value to be established. However, \citet{Downes2007} report that calculating $R_{\rm max}/\upsilon_{\rm max}$ is the best `classical' way of obtaining a single $t_{\rm dyn}$ value (as \citet{Beuther2002}, for example).  Figure \ref{fig6} therefore presents dynamical timescales, calculated via the simplistic $R_{\rm max}/\upsilon_{\rm max}$ for the distance-limited subsample (D$<$6\,kpc), against luminosity, showing no significant correlation (Table \ref{tab5}).

We find that the dynamical timescale dependent parameters $\dot M, \dot P$ and $\dot E$, obtained using the single $t_{\rm dyn}$ = $R_{\rm max}/\upsilon_{\rm max}$, via $\dot M = M/t_{\rm dyn}$, etc., are comparable with those found using the position-variable $t_{{\rm dyn}(x,y)}$, as indicated in Section \ref{lowM}. Furthermore, \citet{Downes2007} note that both these methods can overestimate the flow age and thereby underestimate the timescale-based outflow parameters. The authors find a more accurate representation of $t_{\rm dyn}$ using $1/3\,R_{\rm lobe}/\langle \upsilon \rangle$, as the intensity-weighted-velocity, $\langle \upsilon \rangle$, is probably a better measure of the transverse expansion speed of the lobe and where $R_{\rm lobe}$ is the the perpendicular distance from the jet axis (in their jet driven outflows models). Since the outflows are generally not resolved in this perpendicular direction in our data (and in general single dish studies), this dynamical-time estimate cannot be tested until interferometric observations are obtained.

Interferometric observations have begun to separate some of the complex regions where multiple outflows overlap (e.g. \citealt{Beuther2002c}, Cunningham et al. 2015, submitted to MNRAS), however establishing the source inclination to correct $t_{\rm dyn}$ is still difficult. In light of recent work (\citealt{Peters2014}, \citealt{Klaassen2015}) even outflows resolved on 1-2 arcsecond scales may be a combination of outflows from a small cluster of sources (also see Section \ref{lowM}), thus single outflow $t_{\rm dyn}$ values may not always be measurable.

The major caveat of the dynamical timescale, on which $\dot M, \dot P$ and $\dot E$ depend, is a fairly fundamental one, in that it may not actually be strongly related to the age of the outflows \citep[see,][]{Curtis2010}. $t_{\rm dyn}$ is model-dependent and assumes that material has been accelerated in a shock, probably by a jet, and is travelling outward at a characteristic flow velocity. However, even in the jet scenario, much of the high-velocity material is probably turbulently entrained (at the bow shock or along the jet sides and outflow cavity walls) or accelerated in situ by the passing jet and not have travelled as far as assumed. The outflow may not even trace the jet path well at all, and is certainly slower than the jet. Even in the case of our example source G078.1224$+$03.6320 (IRAS 20126+4104), the driving jet looks as if it has precessed over time \citep{Shepherd2000} whereas the molecular outflow appears to highlight the entirety of its previous and present path. Calculating $t_{\rm dyn}$ using $1/3\,R_{\rm lobe}/\langle \upsilon \rangle$ as reported by \citet{Downes2007} will alleviate some of these issues as this primarily uses outflow parameters at the jet-outflow interaction. This dynamical timescale is independent of the distance travelled by the outflow from the source, however, interferometric observations are a necessity in order to resolve the minor axis of the interaction region. 

Additionally, since the jets associated with outflows are often observed to leave the natal star-forming cores and travel far into the surrounding diffuse ISM \citep{Bally2002,Bally2012}, many molecular (CO) outflows may be `fossils' of a past jet event now coasting under momentum conservation and their size scale may be set by the extent of the core itself. SiO observations could be used as an additional tracer to identify more `active' outflows, as single dish studies detect board wings for many mid-IR bright sources \citep[e.g.][]{Klaassen2007,Lopez2011,Mottram2012}. Recent Sub-Millimeter-Array (SMA) observations by Cunningham et al. (2015, submitted to MNRAS) of NCG 2264 indicate the strong, collimated SiO (5$-$4) outflows are only detected around the IR-dark, most deeply embedded sources in the cluster, which are likely to be the youngest. Such observations would need to be spatially resolved to ensure the SiO emission has the same outflow morphology of the CO emission \citep[e.g.][]{DuarteCabral2014}.

The dynamical timescales should therefore not be over-interpreted as giving accurate source ages or used to calculate accurate accretion rates. They should only be used to estimate dynamic parameters for comparison with other studies following the same methods, and estimating accretion rates to an order of magnitude. Our dynamical timescales are comparable with the deduced accretion timescales for sources covering a range of masses in \citet{DuarteCabral2013}. These authors suggest a roughly constant accretion time for the entire mass range, low- to high-mass. If outflows are intrinsically related to the accretion phase, this would suggest a link between the dynamical timescale of the outflows and accretion timescales for the source. Such a relationship is consistent with what we see in Figure \ref{fig6}, where $t_{\rm dyn}$ is $\sim$constant with source luminosity (i.e. core mass, Paper I). Furthermore, \citet{Mckee2003} find typical timescales for the formation of massive stars as $\sim$\,10$^5$\,yrs, consistent again with our values of $t_{\rm dyn}$, suggesting $t_{\rm dyn}$ may be used as a proxy for accretion timescales. 

That said, it is encouraging to note that the dynamical timescales derived from the outflows compare favourably with the MYSO and H{\sc ii} region phase lifetimes obtained by \citet{Mottram2011b}. These lifetimes represent the total time expected to be spent in each phase, so most sources belonging to either category should be younger than these values. As shown in Figure \ref{fig6} the dynamical times for all sources are equal to or below the phase lifetimes, while also older than the $\sim$10$^{4}$\,yrs predicted by \citet{Davies2011} for sources to be too faint in the mid-IR to be included in the RMS survey.

\subsection{Accretion rates}
\label{accrate}
In principle, if inclination angles were known and dynamical timescales were interpreted as an estimated age of the star cluster driving the outflow and an approximate accretion time, they could then be used to estimate a time-averaged accretion rate $\langle \dot M_{\rm acc} \rangle$ onto the cluster/core (but, crucially, not onto an individual star).

$\langle \dot M_{\rm acc} \rangle$ can be crudely estimated using the average $t_{\rm dyn}$ for each lobe. In Paper I we showed, from fitting in the mass-luminosity plot, that if we assume the luminosity is that of an embedded young cluster, then an average $\sim$40 percent star formation efficiency (SFE) is required in order for the stellar mass of that cluster (with an IMF distribution of stars) to match the measured core mass. Using $t_{dyn}$ = $R_{max}/\upsilon_{max}$, $\langle \dot M_{\rm acc} \rangle = M_{\rm core} \times {\rm SFE} / t_{\rm dyn}$ ranges from $\sim$1.3$\times$10$^{-4}$ to $\sim$8.7$\times$10$^{-3}$\,M$_{\odot}$\,yr$^{-1}$, with a crude average of $\sim$2$\times$10$^{-3}$\,M$_{\odot}$\,yr$^{-1}$, consistent with values used in high-mass star formation models \citep{Mckee2003,Yorke2008,Hosokawa2009,Hosokawa2010}. Note, this is a time averaged accretion rate onto a dense cluster core, containing a distribution of protostars, not onto an individual massive protostar alone. 

The luminosity and accretion rate are correlated at the $P=0.01$ (1 percent, $\sim$2.5\,$\sigma$) level (only using sources D$<$6\,kpc), although this is driven essentially by the core mass itself. Hence the accretion rate and outflow mass are also correlated (but at a less significant level P$\sim$0.05, or 2\,$\sigma$; see Table \ref{tab5}). The main driver being the core mass, which is strongly correlated with the outflow mass, suggests that the most massive cores are indeed accreting more material and therefore the outflow mass itself can be used as a very crude proxy ($\sim$within an order of magnitude) for the accretion rate (under the assumption of t$_{\rm dyn}$ as an order of magnitude age estimate).

Similar arguments, based on the correlation of mass flow rates in outflows with source luminosity and, therefore, protostellar mass, have been used to support the hypothesis of accelerating accretion rates in the formation of massive stars and even to predict birthlines for massive protostars \citep{Norberg2000}. Such arguments require the assumption that the observed molecular-outflow properties of MYSOs can be interpreted as relevant to the sequence of protostellar evolution, rather than as time-integrated or time-averaged quantities produced over the lifetime of the forming star, and may therefore be a logical step too far.

\subsection{Mass Flow, Force and Power}
\label{dyns}
The middle-right and lower panels of Figure \ref{fig5} show the mass flow rate, force (also referred to as momentum flux, e.g., \citealt{DuarteCabral2013}) and power (commonly called mechanical luminosity) of the outflows calculated using the spatially varying dynamical timescale. A clear, linear scaling is seen in these logarithmic plots, albeit with around an order of magnitude scatter, larger than that seen for the $M$, $P$ and $E$ parameters, due to the range of dynamic timescales at a given luminosity (see Section \ref{tdyn}). This is reflected in the correlation coefficients in Table \ref{tab5} which are lower than those for the latter $t_{\rm dyn}$-dependent parameters, albeit still highly significant.
That $\dot M, \dot P$ and $\dot E$ also scale with luminosity is usually taken to indicate a common, scalable driving mechanism (Section \ref{lowM} discusses the relationship with low-mass protostars).

\begin{table*}
\begin{center}
\caption{Dynamic timescale and $t_{\rm dyn}$-dependent parameters calculated for all sources (including those where D$>$6\,kpc) with outflows or with evidence of outflows, where apertures could be defined. The `b', `r', `ave' and `tot' subscripts indicate the blue-shifted lobe, red-shifted lobe, average (of both lobes) and total values of each parameter. $t_{\rm dyn}$ values are in units of 10$^4$\,yr, mass flow rates ($\dot M$) in 10$^{-4}$\,M$_{\odot}$\,yr$^{-1}$, force ($\dot P$) in 10$^{-3}$\,M$_{\odot}$\,km\,s$^{-1}$\,yr$^{-1}$ and power ($\dot E$) in L$_{\odot}$. The full table is available in the online supplementary material. Uncertainties as in Figure \ref{fig5} are 50 percent for $\dot M, \dot P$ and $\dot E$, as discussed in the text, uncertainty in how $t_{\rm dyn}$ is calculated and also due to inclination correction can make these easily an order of magnitude.}
{\footnotesize
\begin{tabular}{@{}lrrrrrrrrrrrrrrrr@{}}
\hline
MSX Source Name && $t_{\rm dyn,b}$ & $t_{\rm dyn,r}$ & $t_{\rm dyn,ave}$ && $\dot{M}_{\rm b}$ & $\dot{M}_{\rm r}$  & $\dot{M}_{\rm tot}$  && $\dot{P}_{\rm b}$  & $\dot{P}_{\rm r}$ & $\dot{P}_{\rm tot}$ && $\dot{E}_{\rm b}$ & $\dot{E}_{\rm r}$ & $\dot{E}_{\rm tot}$ \\
\hline 
  
  G010.8411$-$02.5919  &&    6.0  &    7.7  &    6.8  &&     3.8  &     4.0  &     7.8  &&     1.2  &     1.2  &     2.3  &&  0.4  &  0.3  &  0.7   \\
  G012.0260$-$00.0317  &&   18.6  &   23.6  &   21.1  &&     5.3  &     3.0  &     8.3  &&     2.8  &     1.3  &     4.1  &&  1.6  &  0.6  &  2.2   \\
  G012.9090$-$00.2607  &&    2.5  &    1.9  &    2.2  &&     5.2  &    28.6  &    33.8  &&     3.4  &    26.9  &    30.2  &&  2.4  & 26.0  & 28.3   \\
  G013.6562$-$00.5997  &&    4.3  &    7.2  &    5.7  &&     8.3  &     1.0  &     9.3  &&     4.0  &     0.7  &     4.6  &&  1.9  &  0.5  &  2.4   \\
  G017.6380$+$00.1566  &&    9.8  &    8.0  &    8.9  &&     4.7  &     8.3  &    13.0  &&     1.8  &     4.6  &     6.4  &&  0.7  &  2.4  &  3.1   \\
  G018.3412$+$01.7681  &&    7.7  &    6.7  &    7.2  &&     2.6  &     3.0  &     5.6  &&     0.5  &     1.3  &     1.8  &&  0.1  &  0.6  &  0.7   \\
  G020.7438$-$00.0952  &&   23.2  &   27.7  &   25.4  &&    11.9  &     3.2  &    15.1  &&     5.9  &     1.2  &     7.1  &&  2.7  &  0.4  &  3.1   \\
  G020.7491$-$00.0898  &&   14.7  &   15.9  &   15.3  &&    21.3  &     6.1  &    27.4  &&    11.7  &     2.5  &    14.2  &&  5.8  &  0.9  &  6.7   \\
  G020.7617$-$00.0638  &&   11.3  &    4.6  &    8.0  &&     5.7  &    13.7  &    19.5  &&     2.6  &     7.6  &    10.2  &&  1.1  &  4.7  &  5.8   \\
  G023.3891$+$00.1851  &&    9.9  &    8.9  &    9.4  &&     3.4  &     4.8  &     8.2  &&     1.3  &     2.0  &     3.3  &&  0.6  &  1.0  &  1.6   \\

\hline
\end{tabular}
}
\label{tab4}
\end{center}
\end{table*}

\begin{table}
\begin{center}
\caption{Spearman rank correlation statistics for a range of parameter relationships. `Blue' and `Red' indicate the blue- and red-shifted lobe parameters, respectively; `total' represents the total of both lobes except in the case of $t_{\rm dyn}$ and $\gamma$ correlations, where the average value is used. `$\gamma$$_{\rm thick}$' and `$\gamma$$_{\rm thin}$' are the Mass-spectrum slopes, for which only definite outflow sources with good apertures are used in the correlation. All the correlations where the P-value is $<$0.001 are correlated at least at the 0.001 significance level. The accretion rate is correlated with source luminosity at the quoted significance level, whereas it is only correlated with the outflow mass at the 0.05 level for the quoted $\rho$. P-values of 0.05, 0.002 and $<$0.001 represent the $\sim$2, 3 and $>$3$\sigma$ confidence levels. Size is the value indicating the number of sources in each correlation.}
{\scriptsize
\begin{tabular}{@{}lrrrrrr@{}}
\hline
Correlation with: & Blue  &  & Red & &  Total  &  \\
  (size)          &   $\rho$ & P-value &  $\rho$ & P-value & $\rho$ & P-value \\
\hline
Lum. with (68):  &       & & & & & \\
\hline 
Outflow Mass &      0.57& $<$0.001 &      0.45&  $<$0.001&      0.55&  $<$0.001 \\
Momentum &      0.62&  $<$0.001&      0.53&  $<$0.001&      0.61&  $<$0.001\\
Energy &      0.59&  $<$0.001&      0.54&  $<$0.001&      0.58&  $<$0.001\\
Mass Flow &      0.54&  $<$0.001&      0.50& $<$0.001&      0.55&  $<$0.001\\
Mech. Force &      0.50&  $<$0.001&      0.50&  $<$0.001&      0.51&  $<$0.001\\
Mech. Power. &      0.44&  $<$0.001&      0.45&  $<$0.001&      0.45&  $<$0.001\\
Mean Vel. &      0.29 &    0.02 &      0.34 &   0.005 & 0.30 & 0.12 \\ 
\hline
Lum. with (32):&      & & & & & \\ 
\hline
$\gamma$$_{\rm thick}$  & 0.16&     0.39&     0.25&     0.17&     0.31&    0.09\\
$\gamma$$_{\rm thin}$ & -0.17&     0.35&    -0.27&     0.13&    -0.28&     0.12\\
\hline
Core Mass&        & & & & & \\
 with (48): &        & & & & & \\
\hline 
Outflow Mass   &    0.72&  $<$0.001&      0.64&  $<$0.001&      0.75&  $<$0.001\\
Momentum &      0.77&  $<$0.001&      0.74&  $<$0.001&      0.77&  $<$0.001\\
Energy &      0.70&  $<$0.001&      0.67&  $<$0.001&      0.70&  $<$0.001\\
Mass Flow &      0.65&  $<$0.001&      0.60&  $<$0.001&      0.62&  $<$0.001\\
Mech. Force &     0.57&  $<$0.001&      0.52&  $<$0.001&      0.57&  $<$0.001\\
Mech. Power. &     0.49&  $<$0.001&      0.45&   0.001&      0.49&  $<$0.001\\
\hline 
Accretion &   & & & & & \\
Rate with (48) :&  &   & & & & \\
\hline 
Outflow Mass &  0.33 &  0.02  & 0.30 &  0.04  & 0.37  &  0.01 \\
Source Lum. &  ... & ... & ... & ... & 0.43 &   0.002 \\
\hline 
$t_{dyn}$ with (68):&      & & & & & \\
\hline
Source Lum.  & 0.20  & 0.11 & 0.16 & 0.20 & 0.17  & 0.16\\
Momentum  &  0.17 &   0.15 &   0.24 &  0.05 & 0.19  &  0.11 \\
\hline
Outflow &     & & & & & \\ 
Energy with (48) :&      & & & & & \\
\hline
Turb. Energy &  ...   & ...  & ... & ... & 0.73 & $<$0.001 \\     
Binding Energy & ...   & ...  & ... & ... & 0.71 & $<$0.001 \\     
\hline
\end{tabular}
}
\label{tab5}
\end{center}
\end{table}

\subsection{Low-Mass Analogues and Clustered Sources} 
\label{lowM}
The scaling of outflow properties between low- and high-mass sources could infer a similar driving mechanism, and ultimately a similar star formation scenario for high-mass stars. We compare our results with outflows from single low-mass protostars, known to have jet-driven outflows in their early stages of evolution. Our sample of protostellar outflows are emanating from cores associated with massive star formation \citep{Lumsden2013}. These cores harbour many protostars, some of which are massive, although these cores are at varying stages of evolution as identified by the MYSO or H{\sc ii} region classification. It is not necessarily true that the massive protostars within these cores are responsible for powering the outflows however; i) massive stars may form differently to lower mass stars and not power jets that are thought to drive outflows; ii) what appears to be a massive outflow could be explained by a low-mass protocluster, iii) the massive stars in these cores may be too evolved such that they no longer power the outflows observed. These scenarios are tested below.

\subsubsection{Scaling of outflow force}

\begin{figure*}
\begin{center}
\includegraphics[width=0.90\textwidth]{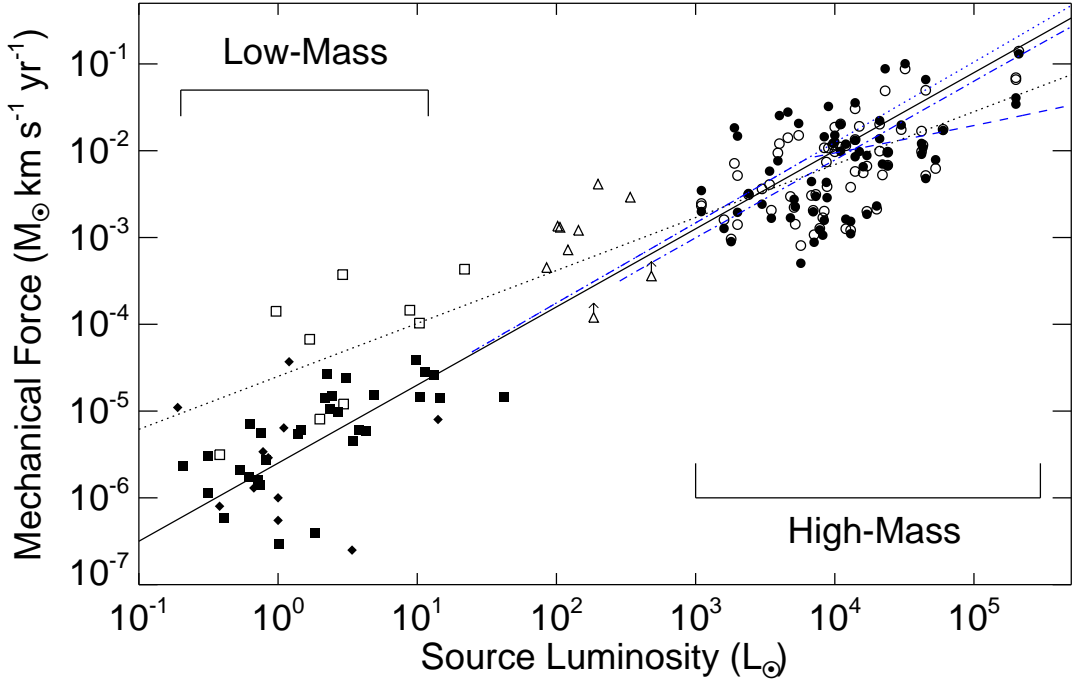} 
\caption{Outflow force $\dot P$ versus luminosity for the distance-limited sample (D$<$6\,kpc). The filled and open circles represent the force calculated with a fixed $t_{dyn}$ = $R_{max}/\upsilon_{max}$ and those summed from the maps where a position variable $t_{dyn(x,y)}$ is used, respectively. Both methods provide comparable parameter values. The open and filled squares are Class 0 and Class I low-mass outflow sources from \citet{Bontemps1996}, the filled diamonds are the $\sim$Class-I sources from \citet{vandermarel2013}, while the open triangles are the high mass outflow Class 0 analogue sources from \citet{DuarteCabral2013}. The black solid line is the linear trend of outflow force with luminosity presented in \citet{Bontemps1996} for their Class I sources, extended to higher luminosities, while the black dotted line is the best fit to our massive outflows extrapolated to lower luminosities. The blue dotted line represent the modelled total outflow force from all protostars in a coeval IMF cluster, the blue dashed is the modelled outflow force if only protostars $M<$8\,M$_{\odot}$ contribute, and the blue dot-dashed line is the modelled outflow force if only the most massive 30 percent of protostars in the cluster provide the outflow (see the text for details).} 
\label{fig7} 
\end{center}
\end{figure*}

Figure \ref{fig7} presents the total outflow force of each outflow source versus luminosity, for our distance-limited sample of massive protostellar cores, together with outflows associated with Class-0 and Class-I, young low-mass YSOs from \citet{Bontemps1996}, $\sim$Class-I, low-mass outflows from \citet{vandermarel2013} (their M7 method in determining force being comparable to ours using $P / t_{\rm dyn}$, where $t_{\rm dyn} = R_{\rm max}/\upsilon$) and the proposed Class-0 analogue, IR-dark high-mass protostars from \citet{DuarteCabral2013}. For consistency, the factor of $\sim$2.9 has been applied to our massive outflows to scale for an average 57.3$^{\circ}$ inclination angle used in \citet{Bontemps1996}. The continuity between the low-mass and high-mass samples is striking. When extended, the best-fit line to the low-mass Class-I sources of \citet{Bontemps1996}, log$_{10}\,F\,= -5.6 + 0.9\times$log$_{10}\,L$(L$_{\odot}$), intersects directly with our massive outflow sample, and, lies slightly below the location of both the low- and proposed high-mass Class-0 analogue sources \citep{DuarteCabral2013}. At face value this suggests, at least for lower mass protostars, a decrease in outflow force with age. When comparing with our more luminous, more massive sources we cannot distinguish between MYSO and H{\sc ii} regions at this resolution (Paper I) to establish if they are at a different evolutionary stages and there are no clear segregations between source types observed. Furthermore, offsets between different data sets may also mimic such a trend dependent on methodology and inclination corrections. The best fit line to our more massive outflows is slightly shallower, Log$_{10}\,F\,=-4.8 + 0.61 \times$ Log$_{10}\,L$(L$_{\odot}$), and extrapolates back to the region between Class-0 and Class-I low mass outflows (Figure \ref{fig7}, black dotted line). At this stage it is unclear whether the shallower slope is due to these cores being protoclusters, rather than a \emph{single} outflow from a single protostar. As previously noted by \citet{Bontemps1996} the slope of all sources in \citet{Cabrit1992} is shallower $\sim$\,0.7, although a linear fit (log-log) to only their sources where $L<$\,100\,L$_{\odot}$ indicates a slope of $\sim$\,1. Independent of the best fit lines, the outflow force is seen to scale over $\sim$6 orders of magnitude in luminosity.

\begin{figure}
\begin{center}
\includegraphics[width=0.45\textwidth]{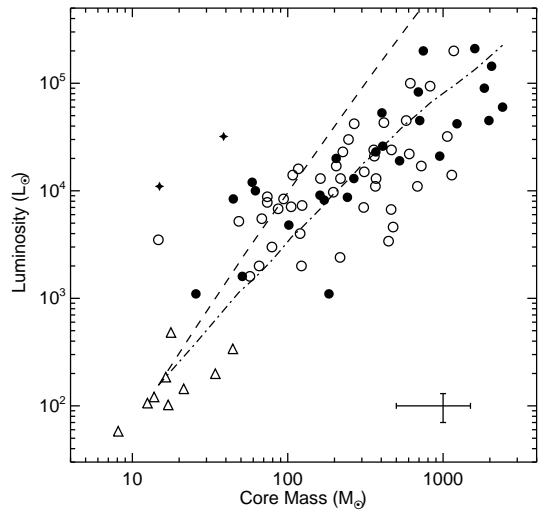} 
\caption{Mass-luminosity plot adapted from Paper I. The open and filled symbols represent MYSOs and H{\sc ii} regions respectively, except the open triangles that are the proposed Class-0 high-mass analogues from \citet{DuarteCabral2013}. The two star symbols represent H{\sc ii} regions in the sample that appear to have dispersed their core material. The dot-dashed and dashed lines represent the luminosity of the most massive star in the IMF cluster and the total luminosity of the cluster respectively (where the star formation efficiency is 50 percent). The \citet{DuarteCabral2013} Class-0 high-mass analogues are mainly positioned below the stellar zero-age-main-sequence (ZAMS) line, which currently intersects our cores. This supports the interpretation that our cores are high-mass Class-0/I analogues in term of position in a mass-luminosity plot.} 
\label{add_ML} 
\end{center}
\end{figure}

In terms of outflow force, our IR-bright MYSOs and compact H{\sc ii} regions are positioned as high-mass analogues of Class-0/I, low-mass YSOs, if simply extrapolating back the fit to our data. A similar correspondence is found if we examine the mass-luminosity relationship as in Paper I (see Figure \ref{add_ML}). The high-mass, Class-0 protostar sample of \citet{DuarteCabral2013} is be positioned below the plotted stellar line as are our cores, albeit with lower luminosities and masses. Note, the Class 0/I classification does not directly indicate a comparable evolutionary stage (or age) for our high-mass sources, as we do not resolve individual protostars within the cores.

The result in Figure \ref{fig7} is very similar to that in \citet{Villiers2014}, who examined a sample of distant outflows associated with methanol-maser sources, which are reliable flags of high-mass star formation \citep{Urquhart2014b}, comparing the outflow force to the clump mass. This is unsurprising, since their clump masses have a near-linear relationship with the embedded MYSO luminosity \citep{Urquhart2014b}, as do our cores (Paper I). However, given that their sources are much more distant, physically larger, more massive and not proven to be a representative sample, we do not undertake a rigorous comparison. Similarly, we acknowledge that there are higher luminosity targets in the sample accumulated by \citet{Wu2004} with which we could also extend our plot, however; some sources are much more distant; the sample is not representative; sources have rough luminosity estimates; and many different methods were used to calculate the outflow parameters.

\subsubsection{Are massive protostars driving the outflows?}
When considering the hypothesis of a scaled outflow driving mechanism, the question arises as to whether there is an upper limit to the YSO luminosity range to which this scaling applies, beyond which there might be a different driving mechanism and, hence, underlying star-formation process. Here we examine whether the most massive stars (M$>$\,8\,M$_{\odot}$, L$>$10$^3$\,L$_{\odot}$) still form via disc accretion and have jets producing outflows, or if the massive outflows from luminous cores are actually a combination of flows from the many low and intermediate-mass sources i.e. can a low/intermediate mass protocluster explain the characteristics of a `single' high-mass outflow. This also simultaneously tests the scenario where the massive protostars in the cluster are more evolved and are no longer powering the outflows. Higher-resolution observations may eventually provide the answer but, meanwhile, we can use the simple model, as used in Paper I, to predict the outflow force from an embedded cluster under the assumptions that the protostars have formed coevally and that force scales with luminosity.

We create a population of protostars following the Salpeter power-law IMF (using only stellar masses from 0.5 to 150 M$_{\odot}$) for a range of model cluster masses and then calculate the corresponding luminosities of the cluster using stellar luminosities from \citet[][Figure 5.11]{Salaris2006} for masses ranging from 0.5\,M$_{\odot}$ to 6\,M$_{\odot}$ and from \citet{Davies2011} for masses $>$6\,M$_{\odot}$. The outflow force for each protostar in the protocluster is calculated according to the relationship for low-mass, nearby, protostars in \citet{Bontemps1996}, log$_{10}\,F\,= -5.6 + 0.9\times$log$_{10}\,L$(L$_{\odot}$), which are more likely to be `single' protostars. The blue lines in Figure \ref{fig7} show the results of the luminosities against the calculated outflow force in each of our test cases. As noted in Paper I, we have not attempted to model a cluster of evolving protostars in detail, given that high mass stars evolve more quickly than lower mass stars, and hence reach their ZAMS luminosity more quickly.

Where the luminosity arises from the complete cluster we find the summed total cluster outflow force is closely consistent with the observations (dotted blue line in Figure \ref{fig7}). All protostars up to $\sim$30\,M$_{\odot}$ must contribute to the outflow force in order to match the observations. The $M\,$=\,30\,M$_{\odot}$ limit is set by the most massive protostar predicted by the model in our most luminous core where D$<$\,6\,kpc. This result supports a similar outflow mechanism for high-mass protostars and thereby supports an upscaled star formation scenario.

We can also consider the case where massive `single' protostars do not form in the same way as their low-mass counterparts, and therefore do not produce outflows at all. This simultaneously tests whether a low/intermediate protocluster can explain the outflow force, and also if the massive protostars have actually stopped powering outflows. We test this by assuming only protostars with masses up to $\sim$8\,M$_{\odot}$, classically low and intermediate mass, contribute to the total outflow force. We find a break in the luminosity-force relation (dashed blue line) at a cluster luminosity of $\sim$6400\,L$_{\odot}$. This is the protocluster luminosity where the most massive protostar reaches $\sim$8\,M$_{\odot}$. This line can be used to highlight two key results. Firstly, the shallower slope of outflow force versus luminosity after $\sim$6400\,L$_{\odot}$ is not consistent with the observations. Thus the scenario where massive stars do not produce outflows either due to a different formation mechanism or being too evolved is unlikely in the case of IMF coeval protoclusters with luminosities over $\sim$6400\,L$_{\odot}$. The second result is that, shortwards of $\sim$6400\,L$_{\odot}$ the line fits almost centrally through the observations and at face value suggest that low/intermediate mass protoclusters can explain the outflow force observed in cores where $L<$6400\,L$_{\odot}$, in cases where higher resolution observations are not available to resolve the sub-structure of the core and identify specific outflow drivers.

The above tests support the interpretation that massive protostars power outflows and form in a similar fashion to low-mass protostars. As noted above, the most massive protostars in a cluster are likely more evolved (closer to ZAMS) than the low/intermediate mass ones. We attempt to account for the non-coeval evolution and test the influence of the most massive protostars only by curtailing the outflow contributing protostars to the most massive 30 percent in the protoclusters. The dot-dashed blue line in Figure \ref{fig7} sits slightly below that extrapolated from the low-mass protostars in \citet{Bontemps1996} but is entirely consistent with the data, suggesting that the most massive protostars in the clusters are responsible for the outflows alone. Recent modelling of outflows from clusters of stars by \citet{Peters2014} do suggest that the most massive protostars dominate the force and power, even when multiple outflows are combined. These authors model four intermediate to massive protostars that form roughly coevally (and have similar masses) from the same accretion disc structure and all produce individual outflows with the same common axis of projection. On the largest scales, at similar resolution to the JCMT observations presented in this paper, a single outflow from the system would be seen. Even their sub-arcsecond, simulated Atacama Large Millimeter/submillimeter Array (ALMA) observations only just begin to separate the bow heads of each outflow, but still cannot disentangle the individual flows.

\subsubsection{The low-mass scaling relationship}

It is possible that the scaling relationship assumed above, from \citet{Bontemps1996} is not representative, being a sample of specifically bipolar low-mass class I sources. Indeed, other samples appear to find shallower slopes e.g. $F\,\propto\,L^{0.7}$, \citep{Cabrit1992}. However, if we repeat our above analysis with this shallower slope (and placing the line such that it still fits the protostars in the lower-mass end, i.e. adjusting the offset), the same key results are obtained. The reported break in luminosity at $\sim$6400\,L$_{\odot}$ does occurs at a lower outflow force, but in general this does not change the conclusion that low/intermediate protoclusters can be responsible for outflows from cores with luminosities below 6400\,L$_{\odot}$. We also reiterate the importance of ensuring that different samples of outflow sources use the same analysis methods in order to be fully compatible and thus remove the ambiguity of estimating offsets to shift the relationships in log-log plots.

Independent of the relationship adopted from low-mass outflows, considering the simplistic model with an ideal, coeval Salpeter IMF distribution of protostars, we find the most likely conclusion is that vast majority of sources in an embedded young cluster contribute to the outflow dynamics and that MYSOs with at least $\sim$30\,M$_{\odot}$ must contribute to the observed flows. The RMS cores are protoclusters and most show a single outflow except for a few cases where multiple outflows are tentatively seen, with non-aligned axes of projection (e.g., G010.8411$-$02.5919 and G109.8715$+$02.1156). In these sources, the outflow parameters of the system are calculated as a whole (as separation of each flow is subjective due to limited spatial resolution) and are indistinguishable from the other `single' outflows we detect.  High-resolution observations do support this as \citet{Klaassen2015} present VLA SiO observations where three massive protostars transitioning to H{\sc ii} regions within a 2000\,au core all have outflows contributing to one large-scale ($>$5000\,au) massive outflow. Our model curtailing the contributing sources to the most massive 30 percent is likely a realistic case for these cores where the massive protostars are closer to the ZAMS and dominate over the outflow dynamics. These conclusions fully support an upscaled outflow mechanism and therefore star formation scenario to massive stars.

\begin{table}
\begin{center}
\caption{Linear fit values for luminosity and core mass relationships. The offset and slope are fits in log-log space and as such correspond to the equation, Log$_{10}$(param) = offset $+$ slope $\times$ Log$_{10}$L(L$_{\odot}$). The (size) indicates how many sources in the correlation. Only sources with distances $<$6\,kpc are used.}
{\scriptsize
\begin{tabular}{@{}lrr@{}}
\hline

Correlation with: & Offset  &  Slope   \\
  (size)          &    &  \\
\hline
Lum. with (68):  &       &  \\

\hline 
Outflow Mass &      -0.67 $\pm$    0.39   &  0.54   $\pm$    0.10     \\
Momentum &   -0.59    $\pm$ 0.42   &  0.67    $\pm$    0.10  \\
Energy &    -0.68   $\pm$  0.52    & 0.78    $\pm$    0.13  \\
Mass Flow &    -5.18   $\pm$  0.34  &   0.49  $\pm$    0.08 \\
Mech. Force &   -5.07  $\pm$   0.46  &   0.61    $\pm$    0.11 \\
Mech. Force inc. corr. &   -4.60  $\pm$   0.46  &   0.61    $\pm$    0.11 \\
Mech. Power. &     -2.92     $\pm$   0.62  &   0.72     $\pm$   0.15   \\ 
Mean Vel. &       0.41     $\pm$ 0.17 &     0.12  $\pm$   0.04 \\
  \\
\hline
Core Mass&          & \\
 with (48): &          & \\
\hline 
Outflow Mass   &    -0.25   $\pm$   0.26   &   0.77  $\pm$  0.11  \\
Momentum &       0.03 $\pm$     0.25   &   0.88   $\pm$  0.11\\
Energy &       0.13   $\pm$   0.33    &  0.98  $\pm$  0.14  \\
Mass Flow &       -4.53   $\pm$   0.23   &   0.55  $\pm$  0.10  \\
Mech. Force &      -4.24 $\pm$     0.32   &   0.67 $\pm$   0.13 \\
Mech. Power. &       -1.93  $\pm$    0.46   &  0.77  $\pm$   0.20 \\
\hline

\end{tabular}
}
\label{tab6}
\end{center}
\end{table}

\subsection{An Alternative Interpretation}
\label{altexp}

\begin{figure}
\begin{center}
\includegraphics[width=0.45\textwidth]{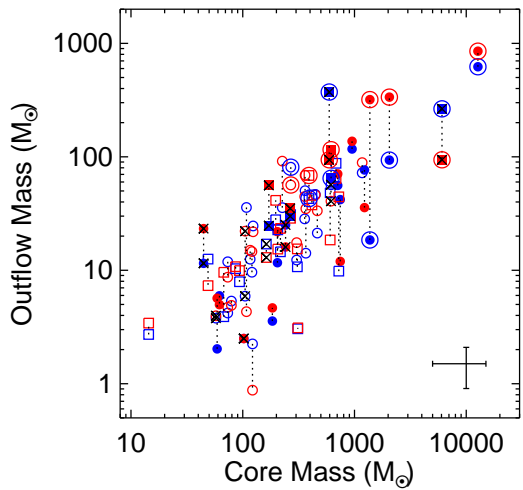} 
\caption{Core mass plotted against the optical-depth-corrected outflow mass for sources whose core masses are flagged as good (see Paper I). There is a clear correlation between the two quantities. The open and filled symbols represent YSOs and H{\sc ii} regions (as Paper I sources that are classified as YSO/H{\sc ii} are identified as H{\sc ii} here), the circles and squares are for `Y' and `S' aperture flags, respectively. The sources circled are those at a distance greater than 6\,kpc while those with a cross symbol are sources with some outflow evidence (flagged `M'). A 50-percent error is plotted for the core mass determination whereas that for outflow masses is the average calculated error as Figure \ref{fig5}.}
\label{fig4} 
\end{center}
\end{figure}

The foregoing discussion explores what might be termed the standard model of bipolar molecular outflows, where we interpret the scaling of outflow properties as an indication that there is a similar jet driven outflow mechanism for high-mass protostars. However, although we see scaling relationships, they may not provide any evidence of a similar outflow driving mechanism, and therefore one cannot assume a scaled up star formation scenario.

An alternative, and considerably simpler explanation for the observed correlations in Figure \ref{fig5} for these high-mass outflows yet to be considered, is that the fundamental relationship is that between core mass and outflow mass, as seen in Figure \ref{fig4} and also reported in \citet{Villiers2014}, and that all the relationships between the $t_{\rm dyn}$-dependent outflow properties and the source luminosity stem from this. There are several pieces of evidence in the data that might point to this model. Firstly, the correlation coefficient for the relationship between outflow mass and core mass, along with that of outflow momentum versus core mass (see Table \ref{tab5}), is the highest of those measured and therefore contains the least scatter, suggesting that one of these two may constitute the basic relationship. 

\begin{figure}
\begin{center}
\includegraphics[width=0.45\textwidth]{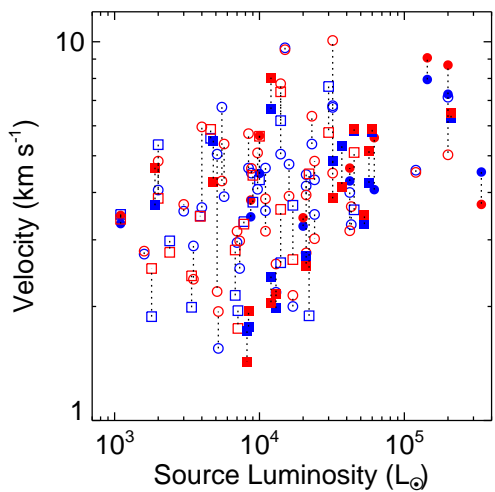} 
\caption{Plot of the average intensity-weighted-mean, `bulk', outflow velocity, calculated via $\langle P_{x,y}/M_{x,y} \rangle$, against source luminosity. There is only a weak correlation over the large range of luminosity, although note each outflow has a range of velocities as used in the calculation of $P$ and $E$. The open and filled symbols represent YSOs and H{\sc ii} regions, whereas the circles and squares are for `Y' and `S' aperture flags respectively.}
\label{fig10} 
\end{center}
\end{figure}

Of the other calculated dynamical outflow parameters, $P$ and $E$ depend on $\upsilon$ and $\upsilon^2$, while $\dot P$ and $\dot E$ are calculated from $M \upsilon /t_{\rm dyn} \sim M \upsilon^2 /R$ and $0.5 M \upsilon^2 /t_{\rm dyn} \sim M \upsilon^3 /R$ and so, in addition to the accelerated mass in the outflow, depend only on various powers of the flow velocity and on the outflow lobe size. Figure \ref{fig6} shows that $t_{dyn}$ = $R_{max}/\upsilon_{max}$ is independent of luminosity, while the evidence for a correlation between the weighted-mean velocity and luminosity is only barely significant in this sample (at the 2$-3 \sigma$ level, see Figure \ref{fig10} and Table \ref{tab5}).  Note that, since the momentum and energy are calculated via the summation over every velocity bin and spatial position, the velocity used is the `bulk' intensity-weighted velocity $\langle \upsilon \rangle$. The slope of a least-squares linear fit (log-log) to the $\langle \upsilon \rangle$ vs $L$ relationship is $\sim$0.12$\pm$0.04, see Table \ref{tab6}. 

With this in mind, the second piece of evidence in the data is that the details of the relationships with luminosity seen in Figure \ref{fig5} are consistent with the way in which the dynamical time based properties are calculated. The slope of the linear fit (log-log) to the dependence of the outflow mass on $L$ is $0.54\pm0.10$, that of the flow momentum is $0.67\pm0.10$ and of flow energy $0.78\pm0.13$ and the relationships involving the mass flow rate, force and power are similar. These slopes differ by only 1$\sigma$ but are consistent with the sequence of increasing powers of $\upsilon$ in their calculation (refer to Table \ref{tab6}). Importantly, we see that the scatter in the relationships also increases in the same sequence. This suggests that the additional powers of $\upsilon$ in calculating $P$ and $E$, etc., is not revealing a more fundamental physical relationship and that $M_{\rm flow}$ has the most basic relationship with $L$, although this, in turn arises because $L$ is linearly dependent on $M_{\rm core}$.

The hypothesis is, therefore, that the entrained mass is the fundamental property of the outflow, that this depends only on the mass available in the dense core, and that the core mass is the only parameter amongst those examined here with a direct physical relationship to the luminosity of the embedded protostars (see Paper I and \citealt{Urquhart2014b}). Thus, all the correlations between the flow, force and power with luminosity could arise simply because there is more mass available to be entrained in massive cores, and these massive cores themselves also tend to contain more luminous protostars, i.e. ranging from single sources, to a cluster, the entrained mass of the outflow is directly related to the mass available to each individual protostar. Such that in the case of a cluster, each protostar drives an outflow related to the protostar's {\em envelope} mass, being a fraction of the total core mass \citep[e.g.][]{Bontemps1996}. Such an interpretation, of course, renders any conclusion regarding the outflow-driving mechanism suspect, since this remains unconstrained and any process that accelerates the core gas with similar entrainment efficiencies will produce essentially the same effect. This, in turn, implies that accretion rates may not be directly inferable from mass outflow rates for high-mass sources as per \citet{Richer2000,Beuther2002,Villiers2014}, as these assume a particular family of momentum-conserving driving mechanisms with particular efficiencies. Indeed, even if more massive protostars drive more powerful outflows, and have higher infall rates (which are intrinsically linked to accretion), one would require knowledge of the efficiencies of the outflow mechanism for high-mass protostars before inferring accretion rates.

There are several \emph{a priori} arguments that may support this hypothesis. Firstly, regardless of how the outflow material is entrained, \citet{Dyson1984} pointed out that the mechanism accelerating the gas of the molecular outflow should be either momentum- or energy-conserving, but not both. Therefore, the virtually equal correlations of outflow force and outflow power with source luminosity always found in studies such as this cannot simultaneously represent a physical relationship. The terminal wind/jet velocity is required as an input for calculation of the outflow force or outflow luminosity in the different regimes \citep[see eqns. 25 and 26,][]{Dyson1984}. Using the wind/jet speed should indicate a constant value of either outflow force or outflow luminosity, while the other parameter scales with source luminosity, dependent on the nature of the outflow (i.e. either momentum- or energy-conserving). 

Secondly, in low-mass sources, the jets are known to leave the densest regions of the molecular cloud \citep[e.g.][]{Bally2002,Bally2012} and travel much further out into the diffuse interstellar environment (often many parsecs but, in any case, on a scale much larger than the molecular outflow), only showing up as bow-shocks when stopped by the diffuse ISM. Thus, an unknown fraction of the jet energy or momentum, depending on what is conserved $-$ most likely momentum $-$ must be deposited in the molecular outflow component and this should create enough scatter in the measured parameters of the latter to wipe out any correlations. Yet low-mass outflow sources produce the most significant correlations \citep[e.g.][]{Bontemps1996}. Whether or not the outflows from high-mass sources are mostly jet-driven is still uncertain (e.g. as the case for G010.8411$-$02.5919, G078.1224+03.6320 and G109.8715+02.1156, see above). 

Furthermore, the parameters of the large-scale Orion molecular outflow \citep[e.g.][]{Erickson1982} are consistent with the standard dynamical relationships \citep{Bally1983b,Cabrit1992} but this outflow is now thought to have been caused by a single impulsive, possibly explosive event \citep{Zapata2009,Bally2011}, unlike the jet-driven flows seen in at least the low-luminosity sources. This alone must raise doubts over the conclusion that the universal correlations imply a single outflow-driving mechanism.

\begin{figure}
\begin{center}
\includegraphics[width=0.45\textwidth]{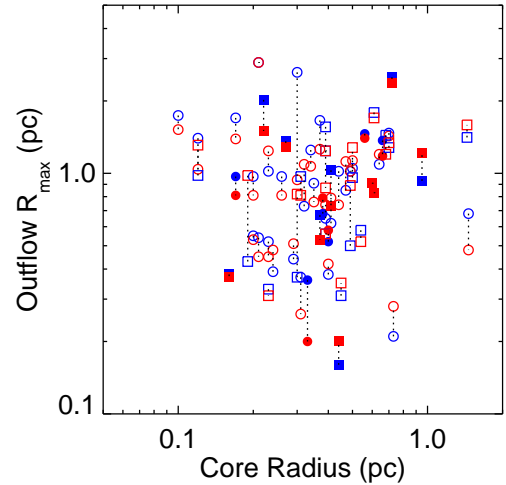} 
\caption{$R_{\rm max}$ for blue- and red-shifted outflow lobes versus the core radius from Paper I.  Open and filled symbols are MYSOs and H{\sc ii} regions, respectively, while the colours represent the blue- and red-shifted lobes, and circles and squares are `Y' and `S' aperture flags. Clearly there is no correlation between these parameters.}
\label{fig_R} 
\end{center}
\end{figure}

Of course, this mass-only model also predicts that the outflow length $R$ will be determined mainly by the density distribution and size of the core. This does not appear to be the case (Figure \ref{fig_R}), however it may be that such a relationship is masked by scatter from the random outflow orientation and by highly variable collimation.

Another consequence of this model is that the dynamical timescales obtained from $R/\upsilon$ cannot be related to real source ages, except in the youngest flows in which a driving jet has not yet reached the edge of the core. In sources in which a jet has left the dense core, $t_{\rm dyn}$ may give only an estimate of the time taken for the jet driven shock to reach the low-density outer regions of the core envelope.  

The power law relationship in Figure \ref{fig4} is $M_{\rm outflow} \propto M_{\rm core}^{0.8 \pm 0.1}$. This is not significantly different enough from a direct linear relationship to warrant too much explanation. If it were, then there would have to be some relationship between the core mass and the fraction of the core affected by the outflow (diminishing if flatter than linear, e.g.) implying some kind of mass-dependent collimation of the flow.

If the preceding model survives further analysis, then the collimated nature of massive protostellar outflows, i.e., their bipolarity, is the remaining feature that unites them to the low-mass version of the phenomenon. This may be due to the ubiquitous presence of disc structures related to accretion. Bipolarity is clear for some of the outflows from this single-dish survey. Some synthetic observations reproduce the bipolar morphology of high-mass outflows, even if driven by ionisation feedback alone \citep[e.g.][]{Peters2012}.  However, the predicted outflow parameter values are lower than those observed and hence magnetic forces are again suggested as an acceleration mechanism. 

The recent review by \citet{Li2014} details that, irrespective of the numerical modelling approach, successful models of jets all include some aspect of a rotating disc and a magnetic collimation and driving force. Only a rare few high-mass sources have been independently studied and indicate a scaled-up picture of low-mass star formation (e.g. Cep A HW 2; \citealt{Patel2005}, rotating material, \citealt{Curiel2006}, proper motion radio jets, and \citealt{Vlemmings2010} outflow and jet aligned magnetic fields). As argued in the review by \citet{Richer2000} a high-resolution, high-sensitivity study of a significant sample of massive protostellar outflows is required. Such a study is still lacking, although the sample in this work provides a firm footing for such an investigation as these high-mass star forming regions clearly drive outflows.

\subsection{The Mass-Velocity Relationship}

\begin{figure*}
\begin{center}
\includegraphics[width=0.95\textwidth]{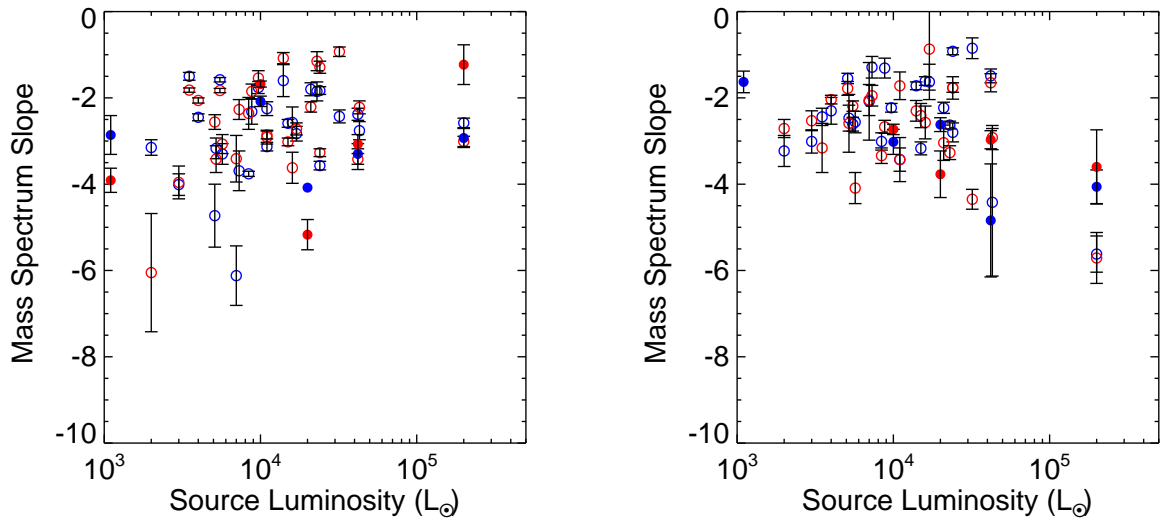} 
\caption{Left and Right: Mass-spectrum slopes for all sources plotted against luminosity for the optical-depth-corrected velocities and the optically thin, higher-velocity ranges in the emission-line wings. The slopes of both velocity regions are comparable. Open and filled symbols are MYSOs and H{\sc ii} regions, respectively, while the colours represent the blue- and red-shifted lobes.}
\label{fig8} 
\end{center}
\end{figure*}

In the earlier stages of massive star formation, the convective surface layers of the protostar may be conducive to the production of magnetic fields which then supply the magneto-hydrodynamic forces required for jet production, powered by the stellar rotation \citep{Hosokawa2010}. Jets, driven and regulated via the complex relationship of MHD forces and disc accretion, may therefore be inherent to massive-star formation \citep[see,][]{Li2014}. 

One method used to investigate jets as the powering mechanism of molecular outflows is to examine the mass spectrum, i.e., the power-law relationship between intensity and velocity \citep{Lada1996,Richer2000,Ridge2001}, in which a break can often be identified in the higher-velocity flow that may be due to molecular dissociation caused by a jet shock \citep{Downes2003,Downes2007}. \citet{Ridge2001} note that optical-depth correction must be undertaken to establish an accurate slope for the relationship, especially at lower velocities. The left and right panels of Figure \ref{fig8} show the results for both optical-depth-corrected and optically thin velocity regimes.

Generally, the slopes are consistent with those found for low-mass outflows \citep{Richer2000,Stojimirovi2006}. However, no clear breaks are seen in the higher-velocity, optically thin regions and single linear relationships fit the mass-spectra of the individual cores reasonably well. The optical-depth-corrected and optically thin slopes also span the same ranges. There is no evidence of a recently accelerated or coasting outflow component \citep{Richer2000}. \citet{Plunkett2015} do not report breaks in the slope of the combined outflows from the two clusters they investigate. This confirms that we could not expect to see a break in any of our high-mass cores as they likely contain multiple, possibly different direction and inclination outflows, which when combined would wash-out any underlying breaks that are seen for single, jet-driven low-mass outflows. The similarity of the slopes for low- and high-mass sources could in itself, be interpreted as due to a similar outflow-acceleration mechanism. Higher sensitivity and higher resolution observations are required to detect the very high velocity flow and confirm its origin from a {\em single} driving source. \citet{Lebron2006} note a steepening mass-spectrum for the high velocity outflow $>$ 40\,km\,s$^{-1}$ in their high-sensitivity single dish observations of G078.1224+03.6320, whereas the JCMT observations here only detect a $\upsilon_{\rm max}$ of $\sim$40\,km\,s$^{-1}$.

\begin{figure*}
\begin{center}
\includegraphics[width=0.95\textwidth]{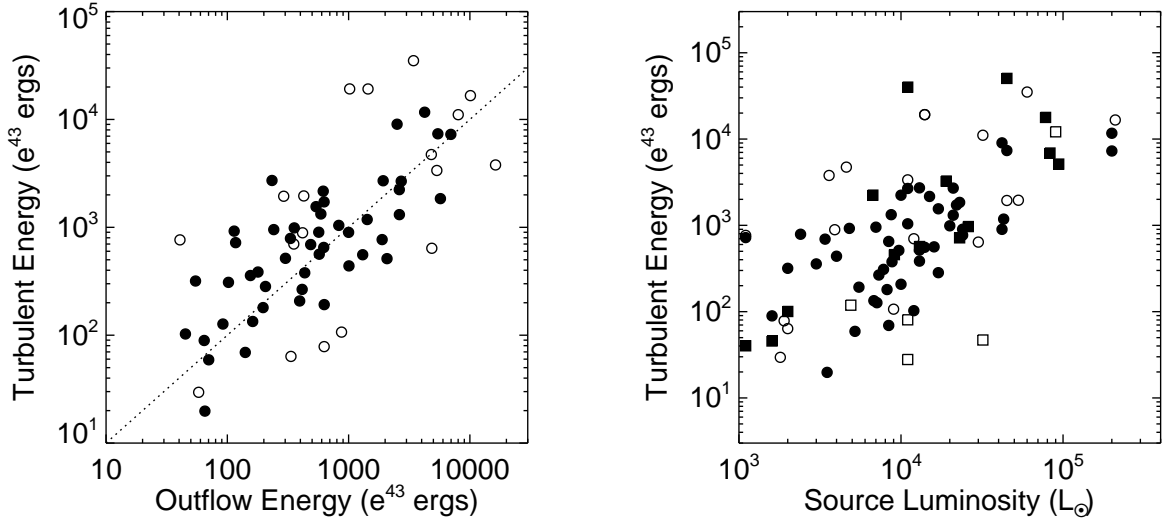} 
\caption{Left: Turbulent energy of the cores plotted against the outflow energy. The open circles represent sources where a core mass is established while the filled ones are subset where core masses are flag 0, 1, and 2 in Paper I, i.e. reliable estimates. On average without inclination correction the outflow energy can fully account for the turbulent energy. Right: Turbulent energy plotted against source luminosity. The open and closed circles are the same as the left figure, the squares are for cores without outflows (open and closed have the same meaning as the circles). It is clear that cores without outflows appear to have the same range of turbulent energy values.}
\label{fig9} 
\end{center}
\end{figure*}

\subsection{Impact on the Natal Core}

Simulations and observations suggest that molecular outflows both can, and cannot have a profound effect on the natal core (e.g. \citealt{Arce2010,Mottram2012,Federrath2014} and also see \citealt{Li2014}). C$^{18}$O data presented in Paper I show 30 sources (D$<$6\,kpc) have smooth velocity gradients across the cores, six of which are aligned with the outflows presented here, suggestive of their strong influence. We can compare the energy content of the outflows to the turbulent kinetic energy in the core gas and the gravitational binding energy of the core (assuming $\rho(r)\,\propto\,r^{-2}$, \citealt{Shepherd2007}). In the case where the thermal motions contribute little to the measured full width half maximum (FWHM) of the core line emission, the turbulent energy can be calculated via E$_{\rm turb}$ = (3/16\,ln\,2)$M_{\rm core}\,\times$FWHM$^2$ \citep{Arce2001}. Figure \ref{fig9} shows the turbulent energy, calculated using $M_{\rm core}$ from Paper I and where the velocity FWHM is measured using C$^{18}$O (3$-$2), versus the outflow energy. For the binding energy calculation, the core radii listed in Paper I are used.

The outflow energy on average can fully account for the turbulent energy of the core and also equates to $\sim$85 percent of the binding energy, even without inclination-angle correction. Hence we might conclude that outflows from massive YSOs contribute significantly to the core turbulent kinetic energy and are a significant source of mechanical feedback. \citet{Cunningham2009} conclude that jet-driven outflows act to maintain the turbulence in a molecular cloud, provided there was already an initial disruption. However, Figure \ref{fig9} (right), shows that the cores that do not contain outflows have turbulent kinetic energies that are consistent with those that do. This suggests that the core turbulence (on scales of $\sim$\,0.4 to $\sim$\,2\,pc) is not driven by the local input from outflows. If the jets that accelerate the molecular flows tend to leave the dense cores, only a fraction of their momentum and energy might be deposited into the dense core gas, and this only in local regions. In the low-mass core, B59, \citet{DuarteCabral2012} see remnant U-shaped cavities and ridges that are a result of the direct impact of outflows on the less dense, local material. However, they also see other velocity structures in the C$^{18}$O data of the denser core, e.g. gradients and infall motions, and although the outflows do have enough energy to fully drive the turbulence, it is not straightforward to conclude that the outflows alone are the predominant production mechanism.

For our cores there are many plausible explications however, simply the outflows are not detected in these sources (e.g. plane of sky confusion, or weak) or there are alternative inputs of turbulence (local H{\sc ii} regions, external winds/shocks, or interactions of the sources within the cores below our resolution). Since turbulent energy tends to flow from larger to smaller scales, we do not expect such localised phenomena to transfer energy to the whole cloud for example. Although, if energy is transferred on core scales at a cavity wall interface, we would require higher spatial resolution to isolate these interaction regions where the cavity is within the beam \citep[e.g.][]{DuarteCabral2012}.

\section{Summary} 
\label{summ}
From a sample of 99 sources, 65 and 20 have been identified to be definitely driving outflows and have some evidence for outflow, respectively. The remaining 14 show no signs of outflows and have Gaussian $^{12}$CO line profiles. For a distance-limited subset (D$<$6\,kpc), 59 have definite outflows, 17 have some evidence of outflows and 13 do not have evidence of outflows.

The kinematic and dynamic parameters have been calculated for all sources with outflows and with evidence for outflows, but only where it was clear how to separate the outflow material from the diffuse ambient emission in integrated maps. Furthermore, $\dot M$, $\dot P$ and $\dot E$ calculated using a position-variable dynamical timescale $t_{{\rm dyn}(x,y)}$ (applied to massive protostellar outflows for the first time) are shown to be consistent with those obtained using a single $t_{\rm dyn}$ estimate via $R_{\rm max}/\upsilon_{\rm max}$. Using these dynamical timescales and SFE values established in Paper I, we infer a time-averaged accretion rate of $\sim$2$\times$10$^{-3}$\,M$_{\odot}$ onto these massive star-forming cores. 

All outflow parameters scale with source luminosity and core mass. There are no intrinsic differences between the cores classified as MYSOs or H{\sc ii} regions. The outflows are driven by many protostars within a cluster (given the resolution of the observations). Specifically, the outflow force scales directly from samples of low-mass protostars, suggesting that our sample consists of high-mass Class 0/I analogues, in terms of position in luminosity-outflow force, and luminosity-mass diagrams. The relationship between outflow force and luminosity is consistent with all the sources in a coeval star-forming cluster contributing to the outflow force. Models in which the massive protostars do not contribute to the outflows are not consistent with the observations, although these models do suggest outflows from low/intermediate mass protoclusters can explain `single' outflows from cores with luminosities $<$6400\,L$_{\odot}$. Models curtailing the mass of the contributing protostars to only the most massive 30 percent are coincident with the observations, suggesting that the massive protostars dominate over the outflow dynamics. The data support a scaling up of the star formation process for massive protostars up to $\sim$30\,M$_{\odot}$ in a star-forming cluster.

An alternative interpretation of the scaling relationships is that the molecular outflow parameters are determined almost entirely by the entrained mass, which is set by the available core mass. This model does not require a single driving mechanism for the {\em molecular} outflow as any form of mass acceleration, explosive or continuous, would produce the observed correlations. The bipolar nature of some of these high-mass outflows may be the key to supporting a scalable driving mechanism from low-mass sources.

Although the outflow energetics are comparable to the turbulent energy in the dense cores, we find that cores with outflows have similar kinetic energies to those without. Such turbulence could be provided by alternative means or by undetected outflows. At the scales probed we cannot establish the local impact these outflows have.

The sample presented is ideal for high-resolution, high sensitivity follow-ups in order to disentangle the smaller groups of protostars driving the outflows, establish their bipolar nature and the underlying driving mechanism.

\section*{Acknowledgments}
We thank the referee for their helpful comments which helped improve the clarity of the paper. Support for this work was in part provided by the Science and Technology Facilities Council (STFC) grant. Work undertaken in this paper made significant use of the \textsc{starlink} software package (http://starlink.eao.hawaii.edu/starlink). This paper made use of information from the Red MSX Source survey database at http://rms.leeds.ac.uk/cgi-bin/public/RMS\_DATABASE.cgi which was constructed with support from the Science and Technology Facilities Council of the UK. The James Clerk Maxwell Telescope has historically been operated by the Joint Astronomy Centre on behalf of the Science and Technology Facilities Council of the United Kingdom, the National Research Council of Canada and the Netherlands Organisation for Scientific Research.

\bibliographystyle{mn2e}

\appendix
\onecolumn
\section{Column density and mass calculation}
\label{AppendixA}

In this appendix the column density and mass equations are derived following from the result of \citet{Garden1991}, except for the CO(3$-$2) transition. The total column density of a linear, rigid rotor molecule under conditions of local thermodynamic equilibrium (LTE), with the populations of all levels characterised by a single excitation temperature, $T_{\rm ex}$, is obtained from the integral of the optical depth over the line profile:

\begin{eqnarray}
\label{eqn_ntot}
N_{\rm tot} = \frac{3k}{8 \pi^{3}B \mu^{2}} \, \frac{{\rm exp}[hBJ(J+1)/kT_{\rm ex}]}{(J+1)} \, \frac{T_{\rm ex} + hB/3k}{[1-{\rm exp}(-h\nu /kT_{\rm ex})]} \int \tau_{\rm \upsilon}\, d\upsilon\; 
\end{eqnarray}

\noindent where $B$ is the rotational constant, $\mu$ is the permanent dipole moment of the molecule and $J$ is the rotational quantum number of the lower state, in this case $J=2$ for the CO(3$-$2) transition. $k$ and $h$ are the Boltzmann and Planck constants respectively. The excitation temperature $T_{ex}$ is solved for, as shown in Equation \ref{eqn1}, Section \ref{optd}.

Here the approximation for $ \int \tau_{\rm \upsilon}\, d\upsilon\ $ follows \citet{Buckle2010} for the case where $\tau \, \ne \,$0:  

\begin{eqnarray}
\label{eqtr}
\int \tau_{\rm \upsilon}\, d\upsilon\, = \left[ \frac{h\nu}{k} \left( \frac{1}{{\rm exp}(h\nu/kT_{\rm ex})-1} - \frac{1}{{\rm exp}(h\nu/kT_{\rm cmb})-1} \right) \right]^{-1} \, \frac{\tau}{[ 1 - {\rm exp} (-\tau)]} \, \int T_{\rm mb}\, d\upsilon\,
\end{eqnarray}

The brightness temperature, $T_{\rm mb}$, is the antenna temperature of the telescope divided by the beam efficiency, T$^{*}_{\rm A}$/$\eta_{\rm mb}$, and corresponds to the Rayleigh-Jeans brightness of a source minus the brightness of the cosmic microwave background with temperature, $T_{\rm cmb}$ = 2.73\,K, over the beam. Combining equations  \ref{eqn_ntot} and \ref{eqtr}, in the limit where $T_{\rm ex}$ $\gg$ $T_{\rm cmb}$ results in the column density:

\begin{eqnarray}
\label{eqn_nave}
N = \frac{3k}{8 \pi^{3}B \mu^{2}} \, \frac{{\rm exp}[hBJ(J+1)/kT_{\rm ex}]}{(J+1)} \, \frac{1}{(h\nu/k)} \, \frac{T_{\rm ex} + hB/3k}{[{\rm exp}(-h\nu /kT_{\rm ex})]} \, \int T_{\rm mb}\, \frac{1}{[ 1 - {\rm exp} (-\tau)]}\, d\upsilon\
\end{eqnarray}

\noindent where the permanent dipole moment for $^{12}$CO is 0.1098 Debye for $^{12}$CO \citep{Chackerian1983}. Conforming to cgs units typically used in such analysis $B$=58.14\,GHz, $k$=1.381$\times$10$^{-16}$\,erg\,K$^{-1}$, h=6.626$\times$10$^{-27}$\,erg\,s, $\nu$($^{12}$CO) = 345.79599\,GHz, velocity $\upsilon$ is in k\,ms$^{-1}$, $\mu$($^{12}$CO) = 0.1098 $\times$10$^{-18}$\,StatC\,cm (where 1\,statC = 1\,g$^{1/2}$\,cm$^{3/2}$\,s$^{−1}$ = 1\,erg$^{1/2}$\,cm$^{1/2}$), $\tau$ becomes $\tau_{12}$, the calculated optical depth of the $^{12}$CO line (Section \ref{optd}) and $T_{\rm ex}$ is the calculated excitation temperature. The column density for the $^{12}$CO (3$-$2) transition is therefore:

\begin{eqnarray}
\label{eqn6_corea}
N({\rm ^{12}CO}) = 4.78\, {\times}\, 10^{12} \, \frac{{\rm exp}(16.74/T_{\rm ex}) \, (T_{\rm ex} + 0.93)}{{\rm exp}(-16.59/T_{\rm ex})} \, \int T_{\rm mb} \, \frac{\tau_{12}}{[1 - {\rm exp} (-\tau_{12})]}\, d\upsilon\;,{\rm cm}^{-2}
\end{eqnarray}

\noindent The mass can then be calculated directly from the column density via:

\begin{eqnarray}
\label{eqn_mass}
M_{\rm gas} = N({\rm CO})\bigg[ \frac{{\rm H_2}}{\rm ^{12}CO} \bigg] \mu_{g}\,m(_{\rm H_2})\Omega\,D^2
\end{eqnarray}

\noindent where $\mu_{g}$ = 1.36 is the total gas mass relative to H$_2$, the abundance ratio ${\rm H_2}/{\rm ^{12}CO}$ = 10$^{4}$, and $D$ is the distance of the source to the Sun, in kpc. $\Omega$ is the solid angle corresponding to the emission in one pixel of the maps used in this work. Thus including the conversion factors the core gas mass in solar masses (M$_{\odot}$) is calculated for every pixel of the outflow lobe maps using Equation \ref{eqn_nave12}. The total outflow lobe masses as reported in Table \ref{tab2} is the summation of the pixel masses within the defined outflow apertures.

\begin{eqnarray}
\label{eqn_nave12} 
M_{\rm gas}\,({\rm M}_{\odot}) = 2.4\, {\times}\, 10^{-12} \, \theta^2 (\arcsec)\, D^2 ({\rm kpc})\, \bigg[ \frac{\rm H_2}{\rm ^{12}CO} \bigg] \, \frac{{\rm exp}(16.74/T_{\rm ex}) \, (T_{\rm ex} + 0.93)}{{\rm exp}(-16.59/T_{\rm ex})}  \, \int T_{\rm mb} \, \frac{\tau_{12}}{[1 - {\rm exp} (-\tau_{12})]}\, d\upsilon\,(cm^{-2})
\end{eqnarray}

\end{document}